\documentclass[11pt]{article} 
\usepackage{longtable}
\usepackage{relsize}
\usepackage{algorithm}
\usepackage{algpseudocode}
\usepackage{amsfonts}
\usepackage{graphicx}
\usepackage{amsmath}
\usepackage{hyperref}
\usepackage{setspace}
\newcommand{\argmin}{\mathop{\mathrm{argmin}}\limits}
\newcommand{\argmax}{\mathop{\mathrm{argmax}}\limits}
\newcommand{\eqdef}{\overset{\Delta}{=\joinrel=}}

\newtheorem{theorem}{Theorem}
\newtheorem{corollary}{Corollary}
\newtheorem{lemma}{Lemma}

\title{A Dichotomous Analysis of Unemployment Benefits\footnote{Citation: Hu, X. (2025). A
Dichotomous Analysis of	Unemployment Benefits. Games, 16(6), 66. \url{https://doi.org/10.3390/g16060066}}
}
\author{Xingwei Hu\footnote{International Monetary Fund, Washington, DC 20431, USA; xhu@imf.org; Tel.: +1-202-492-5271}
\footnote{The views expressed herein are those of the author and should not be attributed to the IMF, its Executive Board, or its management.}
}
\date{}

\begin{document}
\maketitle

\abstract{This paper introduces a novel framework for designing fair and sustainable unemployment benefits, grounded in cooperative game theory and real-time fiscal policy. The labor market is modeled as a coalitional game, where a random subset of participants is employed, generating stochastic economic output. To ensure fairness, we adopt equal employment opportunity as a normative benchmark and propose a dichotomous valuation rule that assigns value to both employed and unemployed participants. Within a continuous-time, balanced budget framework, we derive a closed-form payroll tax rate that is fair, debt-free, and asymptotically risk-free. This tax rule is robust across alternative objectives and promotes employment, productivity, and equality of outcome. The framework naturally extends to other domains involving random bipartitions and shared payoffs, such as voting rights, health insurance, road tolling, and feature selection in machine learning. Our approach offers a transparent, theoretically grounded policy tool for reducing poverty and economic inequality while maintaining fiscal discipline.}




\section{Introduction} \label{sect:intro}
This paper addresses a fundamental issue in economics: not everyone in the labor force can be employed simultaneously. 
Consequently, while some individuals are employed and receive wages along with employment-related benefits---such as pensions, health and life insurance, social security, and paid leave---others remain unemployed. 
This reality raises two key questions: 
\begin{enumerate}
\item Should unemployed individuals receive financial support? 
\item If so, what constitutes a fair amount?
\end{enumerate}

In most advanced economies, the answer to the first question is generally ``yes.'' 
This paper supports that view, advocating for extended unemployment benefits for all unemployed individuals in the labor market, including recent graduates.
In both legal and economic contexts, such financial support may be referred to as unemployment benefits, unemployment insurance, unemployment payments, or unemployment compensation.

To address the second question, this paper proposes a method for fairly distributing unemployment benefits and determining a just payroll tax rate for the employed.
The proposed framework ensures a reasonable allocation of a randomly generated economic output between two randomly formed groups within the labor market. A broader objective is to reduce economic inequality and alleviate poverty. 

The problem of this fair division arises in many real-world contexts. For example, 
consider a $k$-$out$-$of$-$n$ redundant system with $n$ identical components, where the system remains functional as long as any $k$ components are operational. 
Intuitively, each component---whether active or idle---should be valued equally.
A similar principle applies in majority voting systems: even if only some voters support a measure, all voters are assumed to have equal voting power.
In the health insurance industry, not all policyholders fall ill or use their coverage, yet the total cost must be fairly shared among both healthy and sick policyholders.

The labor market presents a more complex version of this familiar problem:
while not everyone can be employed at the same time, everyone desires employment.
Moreover, individuals contribute differently to overall production.
Across these varied scenarios, four common features emerge:
\begin{enumerate}
\item A cooperative group of participants;
\item A random division of these participants into two subgroups;
\item A payoff resulting from this division; 
\item A goal of distributing the payoff fairly among all participants. 
\end{enumerate}

This paper develops a framework to address such situations. 
In systems like the $k$-$out$-$of$-$n$ model or majority voting, the desired outcome is equality. 
The same principle underpins our approach to fair distribution in the labor market, though equality of outcome requires more idealized assumptions.

Determining a fair tax rate and a just distribution of welfare and benefits---both generated by those employed---presents several challenges. 
Although \textit{fairness} is a central concept, it remains inherently abstract and open to multiple interpretations.
This paper argues that prevailing social norms should prioritize equal employment opportunity over equal outcomes or productivity. 

Even when job availability is limited, every individual in the labor market can contribute in some capacity.
Unemployment, therefore, should not be viewed as a personal failure or a systemic flaw.
Instead, it should be understood as a self-regulating mechanism that enhances labor market efficiency, consistent with search and matching models (e.g., \cite{Mortensen&Pissarides1994}).

Applying tax policy to the labor market is further complicated by the economy's dynamic nature and fluctuating productivity levels. For taxation and benefit distribution to be effective, policies must be adaptable to uncertainty and sustainable over time. 
Ideally, these policies should maintain a balance between the value created and the value distributed, even as employment conditions evolve. 

In an objective, data-driven framework, a fair tax rate should be determined solely based on observable data.
This approach reduces the influence of political pressures and minimizes the risk of costly strategic voting. 
It also avoids unnecessary unknown parameters and assumptions, mitigating excessive mathematization, which can lead to misleading conclusions and policy implications (e.g., \cite{Romer2015}).
However, a major challenge is that the heterogeneous-agent production function---which maps all possible combinations of labor participants to economic output---is not observable.
In practice, we only observe the outcomes of specific labor groupings at discrete points in time.
This function is essential for evaluating individual contributions and plays a critical role in assessing benefits and welfare.

Another complication arises from the mismatch in timing between how frequently unemployment data are updated and how often tax rates are adjusted.
Unemployment rates are typically reported frequently (e.g., monthly), whereas tax rates are revised less often (e.g., annually). As a result, policymakers often base tax decisions on outdated unemployment data. 
To address this issue, we propose a fair tax rate model designed for a continuous-time production and employment environment. This model aims to maintain stability in a lower-frequency policy setting by minimizing variability in tax rates.

A substantial body of literature has examined fairness in taxation and unemployment benefits from various perspectives (e.g., \cite{Kornhauser1995}; \cite{Fleurbaey&Maniquet2006}).
A foundational contribution in this area is Shapley's fairness axiom (\cite{Shapley1953}), which led to the development of the Shapley value---a widely used method for allocating employment compensation and welfare (e.g., \cite{Moulin2004}; \cite{Devicienti2010}; \cite{Giorgi&Guandalini2018}; \cite{Krawczyk&Platkowski2018}). 

However, the Shapley value relies on two key assumptions: (1)
all participants directly contribute to value creation, and (2) the distributor possesses complete knowledge of the production function. 
In the context of the labor market, the first assumption is clearly invalid, as the production function does not always increase monotonically with the number of employed individuals. While this assumption may hold within a firm (e.g., \cite{Basu&Mukherjee2025}), it is generally not applicable to broader labor market dynamics.
Furthermore, coalitional games are rarely suited to dynamic contexts---such as those described by Aslam et al. (2024)---due to ever-changing payoff functions, which calls into question the validity of the second assumption.

Recent work by Hu (2006, 2020) relaxes the first assumption and generalizes the Shapley value by introducing non-informative probability distributions to model the division of participants. 
In this framework, a participant is valued either by their marginal contribution---if directly involved in production---or by their opportunity cost to production otherwise. 
Hu (2020) also employs a beta-binomial distribution, among others, to represent equal opportunities for dividing participants.

This paper makes several distinct contributions to the literature on fair and sustainable unemployment benefits. 
First, it extends the dichotomous valuation approach by requiring only a single revealed production outcome,
rendering both marginal contributions and opportunity costs unobservable.
The necessity for complete knowledge of the underlying production function is a key limitation of classical approaches such as the Shapley value. 
While this paper adopts a beta-binomial distribution for the random division of the labor force, many results hold independently of this assumption.

Second, while much of the unemployment insurance literature is empirical or structural, this paper introduces a normative framework grounded in fairness and sustainability.
It provides a game-theoretic microfoundation for financing unemployment insurance, complementing and extending standard macroeconomic models. 
The resulting tax rule is actionable, interpretable, and robust---qualities essential for effective policy design.
This rule relies solely on the unemployment rate and a reserved spending rate for common goods and services. 
Total unemployment compensation is determined entirely by the tax rate and realized production.
Thus, a straightforward calculation can reveal whether a country taxes too little or too much, indicating either a budget imbalance or inequality in employment opportunities.
Additionally, the framework explicitly identifies which production assumptions imply equality of outcome, helping to mitigate income inequality and reduce poverty,
while accommodating disparities in employment opportunities and productivity among labor force participants.

Third, the framework develops a fair-division solution for a random payoff within a coalitional game context.
The proposed method derives a closed-form, continuous-time payroll tax rate that is fair, debt-free, asymptotically risk-free, and devoid of unobservable parameters. 
The solution is robust across alternative objectives and extends naturally to other domains involving random bipartitions and shared payoffs, such as voting rights, health insurance, and road tolling.
For example, when applied to feature selection in machine learning, the fair-division approach achieves significantly higher accuracy than many widely used methods.

Several methodological innovations are introduced or employed in this work.
Policy instruments are systematically linked to Bayesian hyperparameters,
and the balanced budget rule is formulated as a functional equation that holds across all labor market contingencies.
Mechanism design is utilized to back-solve prior hyperparameters---namely, policy variables---to achieve desired posterior properties within the balanced budget framework. Additionally, an identification objective is introduced to bridge gaps between mixed frequencies in the data. 
As a result, the tax solution---expressed as a function of the unemployment rate---is predetermined and transparent, thereby avoiding the human costs associated with frequent rate adjustments.

In deriving this unique and objective solution, the paper deliberately avoids unnecessary complexity, such as randomness, hypotheticals, ambiguity, and latency. 
This includes, but is not limited to, competitive and cooperative labor market dynamics, endogenous job search behavior, non-linear tax schedules, precise labor market size, and fluctuations in the unemployed population. 
While this simplicity entails a degree of abstraction, any real-world application of the fair-division solution should be carefully tailored to its specific context.
Consequently, a fully data-driven approach is adopted, with no unknown parameters to estimate and no residuals to reconcile with a model specification.

The remainder of the paper is structured as follows. 
Section~\ref{sect:D-value} applies the dichotomous valuation (``D-value'') framework developed by Hu (2006, 2020) to evaluate
the value of each individual in the labor market, under the assumption of equal employment opportunity. 
The two components of the D-value are aggregated separately for the unemployed and the employed. 
Section~\ref{sect:bal_identity} introduces a set of fair allocations of net production using these aggregated D-value components and derives corresponding fair tax rates, based on two accounting identities that ensure a balanced budget. 
Section~\ref{sect:tax_rate} identifies a specific tax rate by either maximizing the stability of the posterior unemployment rate or minimizing its expected value. 
This solution is shown to be robust under several alternative criteria. 
Assuming a productivity condition, Section~\ref{sect:LP_EOO} demonstrates that the identified tax rate leads to equality of outcome.
Section~\ref{sect:other_apps} presents an empirical study and explores four additional applications of the division rule: voting power, health insurance, highway tolls, and feature selection. Finally, Section~\ref{sect:conclusion} concludes with policy implications and suggests directions for extending this framework. 
The paper is self-contained, and all mathematical proofs are provided in the appendices.

\section{Dichotomous Valuation}\label{sect:D-value}
Before proceeding with the formal analysis, we introduce some notations.
Consider a general economy in which the labor force comprises both employed and unemployed individuals, with part-time workers excluded for simplicity.
Let $\mathbb{N} = \{1, 2,\cdots,n\}$ denote the set of individuals in the labor force, indexed accordingly.

Let $\mathbf{S} \subseteq \mathbb{N}$ represent the random subset of employed individuals, which varies continuously.
A specific realization of this random subset is denoted by $S$.
For any subset $T \subseteq \mathbb{N}$, let $|T|$ denote its cardinality.
For convenience, we often use lowercase italics to represent cardinalities: $n$ for $|\mathbb{N}|$, $t$ for $|T|$, $z$ for $|Z|$, and $s$ for both $|\mathbf{S}|$ and its realization $|S|$. 

The employment rate is denoted by $\omega=\frac{s}{n} = 1-$ (unemployment rate).
We also use an overbar to denote a singleton set; for example, ``$\overline{i}$'' represents the singleton set $\{ i \}$.
Additionally, we use ``$\setminus$'' for set subtraction, ``$\cup$'' for set union, and $\beta(\theta,\rho)$ for the beta function with parameters $\theta$ and $\rho$.

\subsection{Equality of Employment Opportunity (EEO)}\label{subsect:EEO}

The job market is inherently uncertain, shaped by dynamic interactions among workers, firms, and macroeconomic conditions. Individuals often face incomplete information regarding available opportunities, wages, and future economic trends, while firms must contend with unpredictable demand and technological shifts. This uncertainty is further amplified by global shocks, policy changes, and sectoral transformations, all of which can alter hiring patterns overnight. As a result, job seekers must make decisions under risk---balancing reservation wages, search costs, and expectations---while employers weigh hiring decisions against the possibility of economic downturns and the risk of mismatches with job candidates. These frictions and information gaps create volatility in employment outcomes, rendering the labor market a complex and evolving system rather than a perfectly predictable mechanism.

From a Keynesian perspective, unemployment is often involuntary, arising from insufficient aggregate demand---firms do not hire because they cannot sell their output.
Additionally, labor markets do not clear quickly due to sticky wages and prices.
A moderate degree of unemployment can, in fact, enhance labor market efficiency by mitigating frictions in  search and matching, encouraging skill development, reducing wage-price spirals, facilitating economic adjustment, and motivating firm competition.
Consequently, a new job opening may attract many qualified applicants, and the eventual selection is, to a large extent, determined by chance. 

Describing this uncertainty is prohibitively difficult. 
At the personal level, employment is not a completely random lottery; it is systematically correlated with both observable and unobservable individual characteristics. 
Skills and employment opportunity are generally positively related. 
However, qualified workers may sometimes overlook available opportunities and choose leisure over work at prevailing or lower wages, optimizing their choices intertemporally.
At the sectoral level, the relationship between required skills and opportunities is largely determined by demand and supply, extending beyond individual choice.
Therefore, to model this uncertainty, we relate it to available economic statistics---such as the unemployment rate---and adopt certain simplifying assumptions.

Equal employment opportunity (EEO) is a widely recognized principle and serves as the foundational axiom for our study of fairness.
In the United States, EEO was institutionalized through Executive Order 11246, signed by President Lyndon Johnson in 1965. 
This order prohibits federal contractors from discriminating on the basis of race, sex, creed, religion, color, or national origin.
Similarly, the Equality Act 2010 in the United Kingdom imposes comparable obligations on employers, service providers, and educational institutions.

The literature offers a rich array of both qualitative and formal discussions on the concept of equal opportunity (e.g.,  \cite{Friedman&Friedman1990}; \cite{Roemer1998}; \cite{Rawls1999}).
Given the multiple interpretations of equality, it follows that there are many definitions of fairness---and consequently, numerous possible fair tax rates.

Many scholars interpret fairness or justice as equality of opportunity, especially in the context of distributive justice and economic outcomes. Almås et al. (2023) provides a comprehensive review of fairness preferences and beliefs about inequality, emphasizing that equality of opportunity is a core principle of justice. According to this perspective, differences in outcome are considered unfair if they arise from factors beyond personal responsibility, but fair if they result from individual choices and efforts.

Joseph (1980) examines various conceptions of equality of opportunity, distinguishing it from equality of outcomes. He frames equality of opportunity as a form of procedural fairness, where access to positions is determined by merit and free from discrimination.
In Rawlsian theory (e.g., \cite{Segall2014}), fair equality of opportunity is a foundational element of justice, asserting that individuals with similar talents and willingness to apply them should have equal chances to attain desirable positions, regardless of their social background.

Beyond the justification rooted in equality of opportunity, unemployment compensations can also be supported by other rationales, including social protection, income flow insurance, poverty prevention, and political considerations (e.g., \cite{Sandmo1998}; \cite{Tzannatos&Roddis1998}; \newline \cite{Vodopivec2004}).

In this study, we propose a probabilistic framework for conceptualizing equality of opportunity, wherein employment prospects are presumed to be equitably allocated among all participants in the labor market.
This model is specifically designed to address the principle of EEO, while acknowledging the persistent disparity  between EEO ideals and observed labor market outcomes. 
Nonetheless, policy interventions informed by EEO-based accommodations yield valuable insights for mitigating these inconsistencies. 

Under a distribution-free EEO framework, the probability that any individual 
 $i\in \mathbb{N}$ is employed (i.e., $i \in \mathbf{S}$) is given by $\mathrm{Prob}(i \in \mathbf{S}) = s/n$ for all $i\in \mathbb{N}$.
However, the proportion $s/n$ itself varies over time.
Beta distributions are commonly used to model proportion data (e.g., \cite{Ferrari&Cribari-Neto2004}) due to their continuous range over the interval $(0,1)$, flexible shape, role as conjugate priors in Bayesian analysis, and applicability in regression models with unknown parameters. 

Given that both $n$ (the total labor force) and $s$ (the number of employed individuals) are nonnegative integers, we model the randomness of the employed subset $\mathbf{S}$ using a three-layered uncertainty framework:
\begin{enumerate}
\medskip
\item First Layer: The size of the employed subset $|\mathbf{S}|$ follows a binomial distribution with parameters $(n, p)$, where $p$ is the probability that any given individual is employed.
\medskip
\item Second Layer: The employment probability $p$ is treated as a random variable with a beta prior distribution characterized by hyperparameters $(\theta, \rho)$.
The joint probability density of $p$ and $|\mathbf{S}|=t$ is given by:
\begin{equation} \label{eq:joint_p_t}
\frac{p^{\theta-1}(1-p)^{\rho-1}}{\beta (\theta, \rho)}
\left(\begin{array}{c}n \\ t \end{array} \right)
p^t (1-p)^{n-t} 
=
\left(\begin{array}{c}n \\ t \end{array} \right)
\frac{p^{\theta+t-1}(1-p)^{\rho+n-t-1}}{\beta (\theta, \rho)}.
\end{equation}
This implies that the marginal probability of observing $t$ employed individuals is:
\begin{equation} \label{eq:density}
\resizebox{.75\textwidth}{!}{$
\mathbb{P}(|\mathbf{S}| = t)
=
\mathop{\mathlarger{\int_0^1}}
\left(\begin{array}{c}n \\ t \end{array} \right)
\frac{p^{\theta+t-1}(1-p)^{\rho+n-t-1}}{\beta (\theta, \rho)} \mathrm{d} p  \\
=
\left(\begin{array}{c}n \\ t \end{array} \right)
\frac{\beta (\theta+t, \rho+n-t)}{\beta (\theta, \rho)}
$}
\end{equation}
for any $t=0,1,\cdots,n.$
\medskip
\item Third Layer: 
Given an employment size $t$, all subsets of size $t$ are equally likely to be the employed group $\mathbf{S}$. 
Since there are $\left(\begin{array}{c}n \\ t \end{array} \right)$ such subsets,
the probability of a specific employment scenario $\mathbf{S}=T$ is:  
\begin{equation} \label{eq:probabilitydensity}
\mathbb{P}(\mathbf{S}=T) 
= 
\frac{\mathbb{P}(|\mathbf{S}|=t)}{\left(\begin{array}{c}n \\ t \end{array} \right)} 
= 
\left \{
\begin{array}{ll}
\frac{\beta(\theta+s,\rho+n-s)}{\beta(\theta,\rho)}, \quad&\mathrm{if}\ t=s;\\
0, & \mathrm{otherwise.}
\end{array} 
\right. 
\end{equation}
\end{enumerate}

This three-layered uncertainty model ensures that every individual in the labor market has an equal probability of being employed. 
While this assumption may seem somehow unrealistic---given the realities of structural unemployment and skill mismatches---it is important to clarify that this interpretation of EEO does not imply perfect labor mobility.
Specifically, it does not assume that workers can move freely between jobs, industries, or locations without barriers or costs. 

Furthermore, this EEO does not suggest that all workers are identical; production still requires a heterogeneous workforce.
Nor does it imply that all applicants have an equal chance at any specific job.
Hiring decisions remain at the discretion of employers, who select candidates based on their suitability for particular roles.

EEO is more likely to be realized in labor markets characterized by high market depth, strong labor mobility, low structural unemployment, and minimal skills mismatch.
Diversity and inclusiveness also play a crucial role in enhancing EEO by bridging gaps across different types of positions. 
In such inclusive markets, opportunities are accessible to all---including recent graduates who begin in entry-level roles and build experience toward their ideal careers.

To uphold EEO, governments can adopt several proactive strategies. These include stimulating
job creation across diverse industries to balance labor supply and demand,
providing free training programs for unemployed individuals in high-demand fields,
and developing efficient job-matching platforms to connect job seekers with emerging opportunities---especially as these opportunities shift across sectors.

Female labor participation often lags behind that of males. In response, numerous fiscal policies have been implemented in recent decades to promote greater inclusion.
Professionals can leverage these evolving dynamics to advance their careers or transition into roles that better align with their skills and aspirations.

Since the employment rate $s/n$ results from the interaction between labor demand and supply,
the hyperparameters $(\theta,\rho)$ of the beta distribution can be interpreted as representing the intensity of demand and supply, respectively.
Consequently, $\theta+\rho$ reflects the overall intensity of the labor force---that is, labor force participation.
Explanatory variables may be linked to these intensities.
In particular, $\theta$ and $\rho$ can serve as as policy instruments through which constraints are imposed and objectives are defined.
Their influence is reflected in the posterior employment rate, which emerges from both policy intervention and observed outcomes. 
Based on the profile of this posterior rate, we solve the problem in reverse to determine an optimal policy rule.

According to Equations~(\ref{eq:joint_p_t}) and (\ref{eq:density}), the posterior density function of $p$, given $|\mathbf{S}|=s$,~is:
$$
\frac{\frac{n!}{s!(n-s)!}
	\frac{p^{\theta+s-1}(1-p)^{\rho+n-s-1}}{\beta (\theta, \rho)}
}
{\frac{n!}{s!(n-s)!}
	\frac{\beta (\theta+s, \rho+n-s)}{\beta (\theta, \rho)}
}
= 
\frac{p^{\theta+s-1}(1-p)^{\rho+n-s-1}}{\beta (\theta+s, \rho+n-s)}.
$$
The posterior employment rate thus follows an updated Beta distribution with parameters $(\theta+s, \rho+n-s)$.
We denote this posterior employment rate as $p_{_{n,\omega}}$, given the observation of $n$ and $|\mathbf{S}|=n \omega$. 
In contrast, $p$ remains an unobservable parameter for the binomial random variable $|\mathbf{S}|$.

In practice, equal employment opportunity is frequently hindered by a range of systemic and individual barriers. Organizational practices---such as biased recruitment methods, inflexible job requirements, and limited advancement pathways---often disadvantage underrepresented groups. Cultural factors, including stereotypes and non-inclusive workplace norms, further restrict fair access to opportunities. Inadequate enforcement of EEO laws and insufficient accommodation policies create legal and policy gaps, while implicit bias and disparities in networking at the individual level perpetuate inequality. Additionally, economic and geographic constraints---such as concentrated job markets and unequal access to education---compound these challenges, making it more difficult for all individuals to compete on an equal footing.
Policies derived from EEO, such as those discussed in this paper, help to realize the goals of EEO and overcome these barriers, transforming a hypothetical slogan into concrete reality.

\subsection{Aggregate Values of the Employed and the Unemployed}\label{subsect:Def_D_Value}
For any subset $S\subseteq \mathbb{N}$, we define a heterogeneous-agent production function $v(S)$ to measure the net aggregate production generated by the employed group $S$.
The function $v(S)$ represents net profit, excluding labor costs---which compensate employed individuals for their time and effort in generating $v(S)$.
To isolate the value added by labor, $v(S)$ also excludes the cost of consumed physical and financial resources. 
Both labor and resource costs are exempt from taxation. 
Without loss of generality, we assume $v(\emptyset)=0$ for the empty set $\emptyset$.
Importantly, $v(S)$ does not necessarily increase with the size of $S$; in fact, a certain level of unemployment may enhance productivity. 

To retain labor and minimize turnover, firms often share a portion of their net profit with employees.
For simplicity, we refer to this shared portion of $v(S)$ as \textit{employment welfare}, distinguishing it from \textit{unemployment benefits}, which are provided to those currently unemployed.
Employment welfare plays a critical role in employee retention and business continuation---both of which significantly influence productivity.
The longer employees remain in their roles, the more skilled and efficient they become, and the fewer costly mistakes they make.
In contrast, new hires typically require months to acclimate and years to reach full proficiency.
As key contributors, firm owners also claim a share of $v(S)$ to compensate for their risk-bearing investments and entrepreneurial efforts.

We now formally introduce the components of the D-value. 
For any individual $i \in \mathbb{N}$, we analyze their marginal effect on the value-generation process
by considering two mutually exclusive and jointly exhaustive events:

\begin{itemize}
\medskip
\item Event 1: $i \in {\mathbf S}$ (i.e., the individual is currently employed). 
In this scenario, the marginal contribution of individual $i$ is given by the difference $v({\mathbf S}) - v( {\mathbf S} \setminus \overline{i})$, referred to as the \textit{marginal gain}.
This value represents the additional benefit brought by individual $i$ due to their inclusion in the employed group ${\mathbf S}$. The expected marginal gain is defined as:
\begin{equation} \label{eq:gamma}
\gamma_i[v]
\ \eqdef \
\mathbb{E} \left [v(\mathbf{S}) - v(\mathbf{S} \setminus \overline{i}) \right ]
\end{equation}
where ``$\eqdef$'' denotes definition and ``$\mathbb{E}$'' represents the expectation with respect to the probability distribution of $\mathbf{S}$.
\medskip
\item Event 2: $i \not \in {\mathbf S}$ (i.e., the individual is currently unemployed). 
This may occur if the individual has just entered the labor market (e.g., after completing school) or is experiencing cyclical, structural, or frictional unemployment. 
In this case, the marginal contribution is $v({\mathbf S}\cup \overline{i}) - v({\mathbf S})$, representing the \textit{marginal loss}
due to the absence of $i$ from the workforce.
If $i$ were included in ${\mathbf S}$, production would increase by this amount. 
The expected marginal loss is defined as:
\begin{equation}\label{eq:lambda}
\lambda_i[v] 
\ \eqdef \
\mathbb{E} \left [v({\mathbf S}\cup \overline{i}) - v({\mathbf S}) \right ].
\end{equation}
\end{itemize}

Let $\gamma_i[v]$ denote the employment welfare received by individual $i$ when employed, and $\lambda_i[v]$ the unemployment benefits received when unemployed.
Even if $i$ remains continuously employed, the sets $\mathbf{S}$ and $\mathbf{S}\setminus \overline{i}$ evolve over time, so the
marginal gain is not constant. Similarly, if $i$ remains unemployed, the sets $\mathbf{S}$ and $\mathbf{S}\cup \overline{i}$---and thus the marginal loss---also vary. 
To account for this uncertainty, we define $\gamma_i[v]$ and $\lambda_i[v]$ as expectations, as shown in Equations~(\ref{eq:gamma}) and (\ref{eq:lambda}).

This valuation framework primarily emphasizes the labor force, often overlooking the contributions of other stakeholders. Several implicit considerations underlie this approach:
\begin{itemize}
\medskip
\item Compensation for Human Capital:
In addition to receiving employment welfare, employed individuals are compensated for their labor.
This compensation reflects the value of human capital utilized in generating $v(\mathbf{S})$. 
Human capital is developed not only through current employment but also through prior work experience and pre-employment education. 
In contrast, unemployed individuals receive only unemployment benefits. 
\medskip
\item Nonobservability of Marginal Contributions:
When $i \in \mathbf{S}$, the value $v(\mathbf{S} \setminus\overline{i})$ is unobservable; we can only observe $v( \mathbf{S})$. 
Similarly, when $i \not \in\mathbf{S}$, we cannot simultaneously observe both $v(\mathbf{S} \cup \overline{i})$ and $v( \mathbf{S})$. 
Therefore, it is necessary to transform aggregate marginal values into observable forms, as outlined in Theorem \ref{thm:total_D_value}. 
\medskip
\item Redistribution of Surplus:
The total employment welfare, $\sum\limits_{i\in \mathbf{S}} \left[v(\mathbf{S})-v(\mathbf{S}\setminus\overline{i}) \right]$, does not equal the total value $v(\mathbf{S})$. 
As a result, some of the surplus $v(\mathbf{S}) - \sum\limits_{i\in \mathbf{S}}\left [v(\mathbf{S})-v(\mathbf{S}\setminus\overline{i}) \right ]$ is redistributed to the unemployed individuals in $\mathbb{N}\setminus \mathbf{S}$. This redistribution occurs not through direct transfers, but via government taxation and unemployment benefit systems. 
This mechanism also supports national-level welfare and benefits, as formalized in Theorem \ref{thm:total_D_value}.
\medskip
\end{itemize}

\begin{theorem} \label{thm:total_D_value}
The aggregate components of the D-value are given by:
\begin{equation}\label{eq:aggregated_D_gain}
\resizebox{.8\textwidth}{!}{$
\sum\limits_{i\in \mathbb{N}} \gamma_i [v]
=
\mathbb{E} \left [ \sum\limits_{i\in \mathbf{S}} \left ( v(\mathbf{S}) - v(\mathbf{S}\setminus \overline{i}) \right ) \right ] 
=
\sum\limits_{S\subseteq \mathbb{N}} \frac{s(\theta+\rho-1)-n\theta}{\rho+n-s-1} \
\frac{\beta(\theta+s, \rho+n-s)}{\beta(\theta,\rho)} \ v(S)
$}
\end{equation}
and
\begin{equation}\label{eq:aggregated_D_loss}
\resizebox{.8\textwidth}{!}{$
\sum\limits_{i\in \mathbb{N}} \lambda_i [v] 
=
\mathbb{E} \left [ \sum\limits_{i \in \mathbb{N}\setminus \mathbf{S}} \left (v(\mathbf{S}\cup \overline{i}) - v(\mathbf{S}) \right ) \right ] 
= 
\sum\limits_{S\subseteq \mathbb{N}} \frac{s(\theta+\rho-1)-n(\theta-1)}{\theta+s-1} \
\frac{\beta(\theta+s, \rho+n-s)}{\beta(\theta,\rho)} \ v(S).
$}
\end{equation}
\end{theorem}

\noindent \textbf{Proof}. See Appendix A. \quad $\square$

\medskip
The coalition $\mathbf{S}$ realizes the value $v(\mathbf{S})$, but it could not have achieved this value independently---support from the broader society and the entire labor market is essential.
Accordingly, the government holds the authority to influence aggregate payments, as outlined in Theorem \ref{thm:total_D_value}, by adjusting the hyperparameters $\theta$ and $\rho$.
This control enables the specification of budgetary constraints and the formulation of optimal fiscal policies.

Moreover, substituting an individual in $\mathbf{S}$ with one from $\mathbb{N} \setminus \mathbf{S}$ may yield the same or even greater production value.
Although most jobs are functionally replaceable, neither Equation~(\ref{eq:gamma}) nor Equation~(\ref{eq:lambda}) captures the value added when replacing individual $i$ with another individual $j$ from either $\mathbb{N}\setminus\mathbf{S}$ or $\mathbf{S}$, respectively.

As demonstrated in Theorem 11 of Hu (2020), when $\theta = \rho = 1$ (i.e., when $p$ follows a uniform distribution on $[0,1]$), the Shapley value of player $i$ in the coalitional game $(\mathbb{N}, v)$ equals the sum $\gamma_i [v]+\lambda_i [v]$.
Furthermore, if we impose the functional equation $\sum\limits_{i\in \mathbb{N}} \gamma_i[v] = v(\mathbb{N})$ (or $\sum\limits_{i\in \mathbb{N}} \lambda_i[v] = v(\mathbb{N})$), then $\gamma_i[v]$ (or $\lambda_i[v]$, respectively) also corresponds to player $i$'s Shapley value (see \cite{Hu2020}).
However, these conditions are generally unrealistic in actual labor markets.

\section{Accounting Identities for a Balanced Budget}\label{sect:bal_identity}
According to Equation~(\ref{eq:probabilitydensity}), the expected production is given by:
\begin{equation} \label{eq:totalexpectedvalue}
\mathbb{E} \left [ v(\mathbf{S}) \right ]
= 
\sum\limits_{S\subseteq \mathbb{N}} \frac{\beta (\theta+s, \rho+n-s)}{\beta(\theta,\rho)} v(S)
\end{equation}
where the probability of each employment scenario $\mathbf{S}$ is determined by the size of $\mathbf{S}$ and the hyperparameters $(\theta,\rho)$.
At any given time, only one of the $2^n$ possible employment scenarios is realized.
The specific scenario $\mathbf{S}=S$ occurs with probability 
$$
\frac{\beta (\theta+s, \rho+n-s)}{\beta(\theta,\rho)}
$$ 
and yields a value $v(S)$. 
This is the value available for allocation when scenario $\mathbf{S}=S$ occurs.
In contrast, the Shapley value distributes the total value $v(\mathbb{N})$ among all players in $\mathbb{N}$. Hu (2020) proposes solutions for distributing the expected value $\mathbb{E} \left [ v(\mathbf{S}) \right ]$ among all participants.
Additionally, the literature on unemployment insurance financing (e.g., \cite{Hopenhayn&Nicolini1997}) frequently addresses how to fund such insurance in a sustainable manner.

\subsection{A Real-Time Balanced Budget Rule} \label{subsect:BalancedBudget} 
Our division rule must fully respect all entitlements to $v(S)$.
Each employed individual receives their expected marginal gain, $\gamma_i[v]$, while each unemployed individual receives their expected marginal loss, $\lambda_i[v]$.
Additionally, a portion of $\delta v(S)$ is reserved for common goods and services---supporting the broader economy and society, which in turn sustain the value-generating process.

Thus, the profit-sharing strategy divides the realized net production $v(S)$ into three components:
\begin{enumerate}
\medskip
\item Employment Welfare: 
Based on the coefficients of $v(S)$ in Equations~(\ref{eq:aggregated_D_gain}) and (\ref{eq:totalexpectedvalue}), the employed labor force collectively retains
$$
\frac{s (\theta+\rho-1)-n\theta}{\rho+n-s-1} v(S)
$$ 
as employment welfare. The remaining portion, 
$$
\left [1 - \frac{s(\theta+\rho-1)-n\theta}{\rho+n-s-1} \right ] v(S),
$$
is paid to the government as a payroll tax.
\medskip
\item Unemployment Benefits: 
According to Equations~(\ref{eq:aggregated_D_loss}) and (\ref{eq:totalexpectedvalue}), the unemployed individuals in $\mathbb{N}\setminus S$ collectively receive 
$$
\frac{s(\theta+\rho-1) - n(\theta-1)}{\theta+s-1} v(S)
$$ 
as unemployment benefits.
This allocation does not equal $v(\mathbb{N})-v(S)$, since $v(\mathbb{N})$ is unobservable and may even be less than $v(S)$.
\medskip
\item Public Reserve:
A reserved proportion of $\delta v(S)$ is set aside for collective societal and economic purposes. 
This portion is not distributed directly to individuals.
\medskip
\end{enumerate}

In summary, the allocation rule establishes a functional equation:
$$
v(\mathbf{S}) \equiv \sum\limits_{i\in \mathbf{S}}\left[ v(\mathbf{S})-v(\mathbf{S}\setminus \overline{i})\right]
+ \sum\limits_{i\not \in \mathbf{S}}\left[ v(\mathbf{S}\cup \overline{i})-v(\mathbf{S})\right]
+ \delta v(\mathbf{S}) 
$$
for the random variable $v(\mathbf{S})$, which is a measurable function from $2^\mathbb{N}$ to the real numbers.
This equation holds for any of the $2^n$ possible realization of $\mathbf{S}$ and any value of $v(\mathbf{S})$.
We take expectation on the functional equation.
According to Equations~(\ref{eq:aggregated_D_gain}) and (\ref{eq:aggregated_D_loss}), allocations to $\gamma_i[v]$ and $\lambda_i[v]$ are determined by the formula:
$$
\sum_{i\in \mathbb{N}} \gamma_i[v] + \sum_{i\in \mathbb{N}} \lambda_i[v] = 
(1-\delta) \mathbb{E}\left[ v(\mathbf{S}) \right].
$$
This differs from $(1-\delta) v(\mathbf{S})$. 
Since only one value of $v(\mathbf{S})$ is observed or realized, it is impossible to determine the exact value of $\mathbb{E}\left[ v(\mathbf{S}) \right]$. However, the revealed $v(\mathbf{S})$ serves as both the maximum likelihood and the least squares estimate of $\mathbb{E}\left[ v(\mathbf{S}) \right]$.

We define the tax rate $\tau$ as:
\begin{equation}\label{eq:tax_rate_def}
\tau (\omega,\delta,n) \ \eqdef \ 1 - \frac{n\omega (\theta+\rho-1)-n\theta}{\rho+n-n\omega-1}.
\end{equation}
This rate does not directly depend on the production function $v(\mathbf{S})$.
The tax revenue is used to fund both unemployment benefits and a reserve fund.
Therefore, 
\begin{equation} \label{eq:tax_rate_c} 
\tau (\omega,\delta,n) \equiv \delta + \frac{n\omega (\theta+\rho-1)-n(\theta-1)}{\theta+ n\omega-1}.
\end{equation}

Unemployment benefits may be administered independently or in conjunction with taxation, depending on the country. 
For instance, in Australia, unemployment benefits are integrated into the broader social security system and financed through general taxation.
In contrast, the United States operates a compulsory government insurance system for unemployment benefits, overseeing both the collection and disbursement of funds. 
Alongside the Federal Unemployment Tax Act, individual states manage their own unemployment insurance programs, resulting in variations in tax rates across different years and states.

The reserve component of $v(\mathbf{S})$ serves the public interest rather than individual needs. 
Specifically, it may be allocated to programs and services such as support for individuals outside the labor force, 
Social Security and Medicare, public administration and national defense, public services, and the servicing of past tax deficits, including interest payments.
In the United States, Social Security and Medicare are managed separately and funded through flat tax rates, which are not adjusted based on gross household income---unlike progressive income tax systems.

The continuous-time balanced-budget rule prohibits borrowing or lending across different labor market scenarios and over time. 
This sustainable taxation policy is designed to address the needs of the current market scenario while preserving the ability of future scenarios to meet their own needs. 
Notably, excessive taxation that leads to budget surpluses can also have negative consequences: it may increase consumer prices, reduce output, and suppress competition. Such outcomes can result in a suboptimal allocation of resources, creating deadweight loss and diminishing overall economic efficiency.

Democratic governments frequently encounter challenges in maintaining long-term budget balance.
Successive administrations often lack incentive to address debt inherited from their predecessors.
When GDP is measured using the expenditure approach, governments may opt to finance crises and wars through debt and increased spending, aiming to sustain GDP growth.
Political competition can also result in unsustainable tax cut designed to appeal to specific voter groups.
Over time, these practices contribute to rising debt levels and tax policies that place burdens on future generations.

In contrast, our real-time budget balance framework holds the current administration directly accountable for the debt it incurs.
Nevertheless, counter-cyclical spending---without generating national deficits---remains a crucial instrument for macroeconomic stabilization.
During a crisis, the government can increase the reserve ratio $\delta$ to stimulate public demand, thereby reducing household income.
Unlike the government's one-time distribution of net production, households
can optimize their utility function through intertemporal strategies such as borrowing, saving, lending, and consumption.
Consequently, the government may adopt measures to encourage households to convert savings into spending, or even to finance expenditures through borrowing. For instance, central banks can reduce policy rates, making borrowing more affordable and lowering returns on savings, which discourages saving and promotes spending.
Similarly, temporarily reducing consumption taxes or value-added tax (VAT) can make goods and services more affordable, further incentivizing spending---particularly on durable goods.

As national debt diminishes, the responsibility for debt management increasingly
shifts to households. Individuals generally manage their own debt more effectively than governments, 
primarily due to personal accountability and the immediate consequences of financial decisions. Direct risks---such as damage to credit scores or bankruptcy---motivate individuals 
to borrow prudently and repay debts on time. In contrast, governments often operate 
under political pressures and extended time horizons, favoring short-term expenditures 
to secure voter support while deferring debt obligations to future taxpayers. Unlike 
governments, households cannot print money or issue bonds, which compels them to live 
within their means. Governments' access to these financial tools can foster overconfidence 
and excessive borrowing. Ultimately, the differing incentives and constraints mean that 
private debt management tends to be more transparent and responsible.

In practice, implementing a balanced-budget rule at the level of labor market scenarios---or in real time---is challenging. 
This difficulty arises from the mismatch in frequency between labor market fluctuations and tax rate adjustments.
For example, in the United States, the employment rate $\omega$ fluctuates daily and is reported monthly by the Bureau of Labor Statistics, whereas the tax rate $\tau$ is typically adjusted only once per year. 
To maintain budget balance as effectively as possible, one strategy is to minimize the variance of the employment rate $p_{n,\omega}$ within a given year.
Ideally, this rate would follow a degenerate probability distribution---remaining nearly constant throughout the year.

\subsection{The Set of Feasible Solutions} \label{subsect:FeasibleSet}
For a given triple $(\omega, \delta, n)$, there are three unknowns---$\theta, \rho$, and $\tau$---in a system of two equations, specifically Equations~(\ref{eq:tax_rate_def}) and (\ref{eq:tax_rate_c}). 
Let $\Omega_{\omega,\delta, n}$ denote the set of all feasible combinations $(\theta, \rho, \tau)$ that satisfy both budget constraints:
$$
\Omega_{\omega,\delta,n}
\ \eqdef \
\left \{ 
(\theta,\rho, \tau) \left | 
\begin{array}{l}
\tau \equiv 1 - \frac{n \omega (\theta+\rho-1)-n\theta}{\rho+n- n \omega -1}, \\
\tau \equiv \delta + \frac{n \omega (\theta+\rho-1)-n(\theta-1)}{\theta+ n \omega -1}, \\
0 \le \tau \le 1,\ \theta>0,\ \rho>0.
\end{array}
\right .
\right \}.
$$

There is inherent uncertainty regarding the size of the labor market $n$.
Although $n$ is not treated as a random variable under the EEO assumption, it is considered a time-varying latent variable.
This is due to the lack of a clear boundary between entering and exiting the labor force,  
as well as the fact that many unemployed individuals may not be actively seeking employment. 
Some may also remain unemployed by choice, opting to forgo lower-paid positions. 

Despite its variability and latency, the labor force size $n$ is generally large in macroeconomic contexts. 
Consequently, in addition to solving Equations~(\ref{eq:tax_rate_def}) and (\ref{eq:tax_rate_c}) for specific values of $n$, we aim to derive a tax rule that remains valid for all sufficiently large values of $n$.
We define:
$$
\tau(\omega, \delta) \eqdef \lim\limits_{n\to \infty} \tau(\omega, \delta, n)
$$
if this limit exists for a sequence $(\theta,\rho,\tau)\in \Omega_{\omega, \delta, n}$.
For simplicity, we refer to this limiting value as the "$\phi$-rate," which represents the asymptotic behavior of the tax rate that satisfies the budget constraints.

Equations~(\ref{eq:tax_rate_def}) and (\ref{eq:tax_rate_c}) constitute a system of linear functions in the variables $\theta$ and $\rho$, expressed in terms of $n$, $\omega$, $\tau$, and $\delta$.
For notational convenience, we introduce the following shorthand expressions:
$$
\begin{array}{lcl}
\Delta&\equiv&\omega  +  \tau - \delta \omega  - 1 , \\
\Delta_1&\equiv&\delta \omega - \omega -\tau + 2 \equiv 1 - \Delta >0, \\
\Delta_2&\equiv&\delta \omega \tau - 2 \delta  \omega  - \omega \tau^2 + 2 \omega \tau + \tau -1, \\
\Delta_3&\equiv&(1- \tau)(\delta-\tau) = \delta -\tau -\delta\tau + \tau^2 < 0, \\
\Delta_4&\equiv&- \delta \omega \tau + \delta  \omega + \delta \tau  - 2\delta +  \omega\tau^2 - 2 \omega\tau + \omega  -  \tau^2 + 3\tau -1.
\end{array}
$$
All these expressions are bounded since $0\le \omega, \tau, \delta\le 1$. 
Lemma~\ref{lm:solvethetarho_delta}, which is used in the proofs of subsequent theorems, provides a method for extracting $\theta$ and $\rho$ from \mbox{Equations~(\ref{eq:tax_rate_def}) and (\ref{eq:tax_rate_c}).}

\begin{lemma} \label{lm:solvethetarho_delta}
In terms of  $(n,\delta,\tau, \omega)$,
$\theta$ and $\rho$ can be extracted from Equation~(\ref{eq:tax_rate_def}) and Equation~(\ref{eq:tax_rate_c})~as:
\begin{equation}\label{eq:theta_rho_in_tau} 
\left \{
\begin{array}{rcl}
\theta&=&\frac{n^2 \omega \Delta_1 + n\Delta_2 +\Delta_3 }{n\Delta + \Delta_3}, \\
\rho  &=&\frac{n^2 (1-\omega) \Delta_1 + n\Delta_4 + \Delta_3}{n\Delta + \Delta_3}.
\end{array}
\right .
\end{equation}
\end{lemma}

\noindent \textbf{Proof}. See Appendix B. \quad $\square$

\medskip
Given that $\Delta_1>0$ and $\Delta_3<0$ in Equation~(\ref{eq:theta_rho_in_tau}), we require $\Delta>0$ to ensure that both $\theta$ and $\rho$ remain positive for large but finite values of $n$.
Consequently, Corollary~\ref{cr:minimal_phi_rate} states that a government should levy at least $1 - \omega + \delta \omega$ of the production as payroll tax.

\begin{corollary} \label{cr:minimal_phi_rate}
The minimal $\phi$-rate is $\tau(\omega,\delta) = 1 - \omega + \delta \omega$.
\end{corollary}

The posterior employment rate, $p_{_{n,\omega}}$, reflects a balance between observed data and policy-specified prior expectations. 
As the volume of data increases alone, the influence of the prior or policy diminishes.
Accordingly, Theorem \ref{thm:limit_distribution} states that $p_{_{n,\omega}}$ converges to a degenerate distribution as $n\to \infty$, provided that $\Delta>0$.
This implies that a fixed tax rule does not affect the central tendency of $p_{_{n,\omega}}$, as long as the tax rate satisfies $\tau(\omega,\delta)>1-\omega+\delta \omega$.

\begin{theorem}\label{thm:limit_distribution}
For any $\phi$-rate $\tau \in (1-\omega+\delta\omega,1)$, as $n\to\infty$,
the posterior employment rate $p_{_{n,\omega}}$ converges in distribution to a degenerate probability distribution concentrated at $\omega$.
\end{theorem}

\noindent \textbf{Proof}. See Appendix C. \quad $\square$

\medskip
This result motivates further exploration of additional properties of $p_{_{n,\omega}}$, such as its variance, rate of convergence, and behavior under varying tax rates---especially given that the labor force size
$n$ is finite, regardless of its magnitude.
Thus, holding $n$ finite, we set objectives for the properties of $p_{_{n,\omega}}$, and then back-solve the policy variables $(\theta, \rho)$. In other words, we design policies so that the employment rate $p_{_{n,\omega}}$ exhibits certain desirable characteristics.

To derive a unique solution from the set $\Omega_{\omega,\delta,n}$ or its limiting boundary, an additional constraint or objective must be imposed, 
as infinitely many solutions exist for a given triple $(\omega,\delta, n)$. 
One approach is to leverage the statistical relationship between $p$ and the hyperparameters $(\theta, \rho)$: the mean and mode of $p$ are $\frac{\theta}{\theta+\rho}$ and $\frac{\theta-1}{\theta+\rho-2}$, 
respectively (e.g., \cite{Johnson&Kotz&Balakrishnan1995}, Chapter 21). 
Similarly, the mean and mode of $p_{_{n,\omega}}$ are $\frac{\theta+s}{\theta+\rho+n}$ and $\frac{\theta+s-1}{\theta+\rho+n-2}$, respectively.

For example, one might set either the mean or the mode equal to the average employment rate from the previous year. 
Alternatively, these expressions could be aligned with a target or natural employment rate. 
However, this identification strategy requires additional input:
historical averages may not reflect recent trends, natural rates are prone to estimation errors, and target rates may be unattainable. 

An alternative approach involves determining a tax rate that maximizes certain properties of $p_{_{n,\omega}}$, such as its expected value for large but finite values of $n$. 
If the $\phi$-rate depends on the realized value of $p$, rather than its unobservable uncertainty, a natural way to eliminate this uncertainty is to minimize it or analyze the posterior rate $p_{_{n,\omega}}$, treating $\omega$ as fixed while allowing $\theta$ and $\rho$ to remain at the decision-maker's discretion. 

While $\omega$ provides informative insight into the central tendency of the posterior labor market as $n\to\infty$, 
other critical aspects of the full profile of $p_{_{n,\omega}}$ include its asymptotic dispersion and the labor market's response to the $\phi$-rate $\tau(\omega, \delta)$---especially when $n$ is large but finite.
Therefore, optimal criteria over the domain $\Omega_{\omega,\delta, \infty}$ can still be established by minimizing dispersion or optimizing the expected market response.
For instance, selecting the $\phi$-rate that minimizes dispersion results in the fastest convergence to the degenerate distribution described in Theorem \ref{thm:limit_distribution}.
This approach helps bridge the frequency gap between the daily employment rate $\omega$ and the annually adjusted tax rate $\tau$.

\section{An Optimal Fair Tax Rate}\label{sect:tax_rate}

In this section and the next, we derive the minimal $\phi$-rate, $\tau(\omega, \delta) =1-\omega+\delta\omega$, from multiple perspectives. 
A well-designed distribution of benefits should not compromise market efficiency, and an effective tax policy must  preserve incentives for employment and productivity, as emphasized in Theorems \ref{thm:max_posterior_mean} and \ref{thm:compare_gamma_lambda}. 
At the same time, a robust and optimal tax rule should satisfy several key criteria.
Accordingly, we examine five such criteria in Theorems \ref{thm:zero_variance}, \ref{thm:semivariance}, \ref{thm:max_posterior_mean}, \ref{thm:MAD_mean}, and \ref{thm:equality_of_outcome}---each of which independently identifies the same solution. 
Collectively, these criteria aim either to minimize employment market risk or to maximize expected employment, all while respecting the constraints of market capacity and budget~balance.

The formulation of this tax rule is grounded in observed market behavior. 
While an efficient labor market enhances productivity $v(\mathbf{S})$, a higher employment rate does not necessarily imply higher productivity---and vice versa. 
Acemoglu and Shimer (2000) argue that a moderate level of unemployment can actually improve productivity by enhancing  job quality.
Unemployment can foster peer pressure to increase output, facilitate the reallocation of labor from declining firms, and enable growing companies---and the broader economy---to respond more effectively to external shocks. 
Therefore, our proposed tax rule does not aim merely to maximize the employment rate. Instead, it is rooted in empirical  market behavior and is designed to enable the market to achieve the highest feasible employment rate within the constraints of budgetary discipline and market capacity.

\subsection{Asymptotic Risk-Free Tax Rate}
A stable tax rate creates a favorable environment for maintaining budget balance, reducing labor market uncertainty, and encouraging investment in both technology and human capital. 
When viewed as a function of the employment rate $\omega$, the tax rate $\tau$ acts as a mechanism for transmitting risk from fluctuations in unemployment to fiscal policy.

In the absence of external shocks, a stable tax rate effectively corresponds to a stable unemployment rate.
However, directly targeting a stable employment rate $\omega$ does not eliminate its variability, as it remains susceptible to numerous shocks.
Instead, a well-designed tax policy can mitigate the magnitude of these risks and address inequities arising from the mismatch in timing between the frequently changing employment rate $\omega$ and the typically annual adjustment of the tax rate $\tau$.
For example, if $\omega$ is high in the first half of the year and low in the second, a constant tax rate throughout the year ensures equal payment to unemployed individuals in both periods.
However, under a tax rule such as the $\phi$-rate $\tau(\omega, \delta) = 1 - \omega + \delta \omega$, unemployment benefits may still differ between periods.

\begin{theorem}\label{thm:zero_variance}
The $\phi$-rate $\tau = 1 - \omega + \delta \omega$
minimizes the asymptotic variance of the posterior employment rate $p_{_{n,\omega}}$, that is:
$$
\argmin_{\tau} \lim\limits_{n\to\infty} \mathrm{VAR} \left (\sqrt{n} p_{_{n,\omega}} \middle| (\theta,\rho,\tau) \in \Omega_{\omega,\delta,n}\right ) 
= 
1 -\omega +\delta\omega,
$$
and the minimum limiting variance is zero.
\end{theorem}

\noindent \textbf{Proof}. See Appendix D. \quad $\square$

\medskip
When the variance of $\sqrt{n} p_{_{n,\omega}}$ is used as a measure of instability, Theorem \ref{thm:zero_variance} demonstrates that the $\phi$-rate tax rule $\tau(\omega, \delta) = 1 - \omega + \delta \omega$ minimizes this asymptotic variance. 
However, several clarifications are necessary:
\begin{enumerate}
\medskip
\item Limit Behavior:
The stability described is achieved only in the limit as $n \to \infty$, where the variance approaches zero. 
In practice, however, $p_{_{n,\omega}}$ remains subject to exogenous shocks, as discussed by Pissarides (1992) and Blanchard (2000). Consequently, when these shocks are taken into account, the minimum limiting variance is greater than zero.
\medskip
\item Practical Adjustment: While $1-\omega+\delta\omega$ is the limiting tax rule, 
for large but finite $n$, a small positive adjustment may be added to ensure the positivity of $\theta$ and $\rho$.
For~example, 
$$
\tau (\omega,\delta,n)=1-\omega+\delta\omega+\frac{2\omega(1-\omega)(1-\delta)^2}{n}
$$ 
ensures that the denominators in Equation~(\ref{eq:theta_rho_in_tau}) remain positive, thereby guaranteeing $\theta>0$ and $\rho>0$. 
This adjustment becomes negligible as $n$ grows large.
\medskip
\item Labor Mobility: With near-zero variance in the unemployment rate, labor mobility implies that layoffs and  new hires nearly offset each other, keeping the total employment size $s$ nearly constant. 
It also means that the number of employed individuals $s$ changes proportionally with the labor market size $n$.
Consequently, the employment rate $s/n$ remains stable.
\medskip
\item Asymmetric Risk Minimization: Although the posterior distribution is skewed, the $\phi$-rate tax rule minimizes both total and one-sided risks in $\sqrt{n} p_{_{n,\omega}}$, as further elaborated in Theorem \ref{thm:semivariance}. 
Policymakers are particularly concerned with downside risk.
\medskip
\end{enumerate} 

\begin{theorem}\label{thm:semivariance}
As $n\to \infty$, the $\phi$-rate $\tau(\omega, \delta) = 1 - \omega + \delta \omega$
minimizes both the lower and upper semivariance of $\sqrt{n} p_{_{n,\omega}}$ where $(\theta,\rho,\tau)\in \Omega_{\omega,\delta,n}$.
\end{theorem}

\noindent \textbf{Proof}. See Appendix E. \quad $\square$

\medskip
\subsection{Consistency and Robustness}

The $\phi$-rate $\tau(\omega, \delta) = 1 - \omega + \delta \omega$ encapsulates several essential features of the labor market. 
First, it represents the policymaker's optimal response for stimulating employment while adhering to the constraints outlined in Equations~(\ref{eq:tax_rate_def}) and (\ref{eq:tax_rate_c}).
Traditional unemployment insurance literature (e.g., \cite{Baily1978}; \cite{Chetty2006}) emphasizes the trade-off between providing income smoothing for the unemployed and maintaining incentives to search for work.
Second, this tax rule can also be derived by minimizing statistical dispersion measures that are more robust than variance or semivariance. 
Additionally, it contributes to reducing income inequality.

Importantly, the tax policy is predetermined before any specific economic scenario $\mathbf{S}$ unfolds. 
This transparency reduces both the size and influence of government, which might otherwise spend months deliberating over tax schedules. 
It also prevents opportunistic behavior, such as favoring certain employment scenarios or manipulating unfavorable ones to justify debt.

The rule $\tau(\omega, \delta) = 1 - \omega + \delta \omega$ represents an effective taxation strategy that maximizes
employment without compromising equal opportunity or budget balance. 
For policymakers, a central concern is the forward-looking employment profile $p_{_{n,\omega}}$. 
Unlike Theorem~\ref{thm:limit_distribution}, which considers the limit as $n\to \infty$ for a fixed $\tau$, 
Theorem~\ref{thm:max_posterior_mean} holds $n$ fixed and allows $\tau$ to vary.
According to Theorem \ref{thm:max_posterior_mean}, the mean of $p_{_{n,\omega}}$ decreases as $\tau$ increases, for any fixed, large $n$. 
Therefore, to maximize the posterior mean, the tax rate should be minimized---while still satisfying the condition $\tau(\omega, \delta) \in [1-\omega+\delta\omega,1]$.
Thus, the $\phi$-rate that maximizes the mean of $p_{_{n,\omega}}$ is $\tau(\omega, \delta) = 1-\omega+\delta\omega.$ 
The condition $\omega >0.5$ in Theorem~\ref{thm:max_posterior_mean} is typically met in real-world economies.

\begin{theorem}\label{thm:max_posterior_mean}
For any $\omega \in (0.5,1)$ and a finitely large $n$, the mean  of $p_{_{n,\omega}}$ decreases as the $\phi$-rate $\tau \in (1-\omega+\delta\omega,1)$ increases. 
Consequently,
$$ 
\lim\limits_{n\to\infty} \argmax_{\tau} \ \mathbb{E}\left[p_{_{n,\omega}} \bigm\vert (\theta,\rho, \tau)\in \Omega_{\omega,\delta,n}  \right] 
=
1 -\omega +\delta\omega.
$$
\end{theorem}

\noindent \textbf{Proof}. See Appendix F. \quad $\square$

\medskip
Moreover, this tax rule also minimizes the mean absolute deviation (MAD) from the mean as $n \to \infty$.
For Beta distributions---especially with large parameters---MAD is a more robust measure of dispersion than variance or semivariance. 

The MAD for the posterior $p_{_{n,\omega}}$ is given by (e.g., \cite{Gupta&Nadarajah2004}, p. 37):
\begin{equation}\label{eq:MAD_mean}
\mathbb{E} \left [ \left |p_{_{n,\omega}} - \mathbb{E} (p_{_{n,\omega}}) \right |  \right ]
=
\frac{2(\theta+s)^{\theta+s} (\rho+n-s)^{\rho+n-s}}{\beta (\theta+s, \rho+n-s)(\theta+\rho+n)^{\theta+\rho+n}}.
\end{equation}

\begin{theorem}\label{thm:MAD_mean}
The $\phi$-rate $\tau(\omega, \delta) = 1 - \omega + \delta \omega$ minimizes the asymptotic MAD of $p_{_{n,\omega}}$, that is:
$$
\argmin_{\tau} \lim\limits_{n\to\infty} 
\mathbb{E} \left [ \left| n p_{_{n,\omega}} - \mathbb{E} (n p_{_{n,\omega}}) \right| \biggm\vert (\theta,\rho, \tau)\in \Omega_{\omega,\delta,n} \right ] 
= 1 -\omega +\delta\omega.
$$
\end{theorem}

\noindent \textbf{Proof}. See Appendix G. \quad $\square$

\section{Labor Productivity and Equality of Outcome}\label{sect:LP_EOO}

An individual's productivity is shaped by their workplace environment. 
For any individual $i\in \mathbf{S}$, their productivity within $v(\mathbf{S})$ reflects how consistently and efficiently they complete tasks and achieve goals. 
However, actual performance---measured as $v(\mathbf{S})-v(\mathbf{S}\setminus \overline{i})$---depends on the specific composition of the group $\mathbf{S}$.
Even with a fixed group size $s$, the production $v(\mathbf{S})$ varies depending on the members of $\mathbf{S}$.
When comparing two individuals, one may outperform the other in certain scenarios but not in others.

At the macroeconomic level, individual productivity does not always align with national productivity, since the marginal contribution $v(\mathbf{S})-v(\mathbf{S}\setminus \overline{i})$ does not perfectly correlate with total output $v(\mathbf{S})$. 
However, under the proposed distribution rule for $v(\mathbf{S})$, aggregate employment welfare aligns exactly with total production for a given tax rate $\tau$.
This alignment incentivizes labor market efficiency.
Meanwhile, the imperfect alignment of individual incentives fosters labor mobility, which also enhances overall efficiency.

To connect personal productivity with incentives, we introduce a partial ordering in the labor market.
For any two individuals $i, j\in \mathbb{N}$ with $i\not = j$, we say that individual $i$ \textit{uniformly outperforms} individual $j$ in the production function $v$ if the following two conditions hold:

\begin{itemize}
\medskip
\item $v(T\cup \overline{i}) - v(T) \ge v(T\cup \overline{j}) - v(T)$ for any $T\subseteq \mathbb{N} \setminus \{i, j\}$;
\medskip
\item $v(T)-v(T\setminus \overline{i}) \ge v(T)-v(T \setminus \overline{j})$ for any $T\subseteq \mathbb{N}$ with $\{i,j\} \subseteq T$.
\medskip
\end{itemize}

\noindent These conditions imply that individual $i$ has higher marginal productivity than individual $j$ in all comparable employment scenarios---whether both are employed or both are unemployed. 

Since productivity plays a central role in Equations~(\ref{eq:gamma}) and (\ref{eq:lambda}), individual $j$ should receive less employment welfare and fewer unemployment benefits than individual $i$. 
This principle is formalized in Theorem \ref{thm:compare_gamma_lambda}. 
Importantly, the theorem does not rely on the beta-binomial distribution from Equation~(\ref{eq:probabilitydensity}), provided that individuals $i$ and $j$ have equal employment opportunity. Other individuals in the labor market may still face unequal opportunities.
Moreover, the theorem holds for all tax rates, including the specific $\phi$-rate: $\tau(\omega, \delta)=1-\omega+\delta\omega$.

\begin{theorem}\label{thm:compare_gamma_lambda}
If individual $i\in \mathbb{N}$ uniformly outperforms individual $j\in \mathbb{N}$ in production $v$, and both have equal employment opportunity, then $\gamma_i[v] \ge \gamma_j[v]$
and $\lambda_i[v] \ge \lambda_j[v]$.
\end{theorem}

\noindent \textbf{Proof}. See Appendix H. \quad $\square$

\medskip
We define individuals $i$ and $j$ as \textit{symmetric} in the production function $v$ if each uniformly outperforms the other. 
According to Theorem \ref{thm:compare_gamma_lambda}, when $i$ and $j$ are symmetric and have equal employment opportunities, they should receive identical unemployment benefits when both are unemployed, and equal employment welfare when both are employed.

Although the assumption of bilateral symmetry or uniform outperformance is conceptually appealing, it is often overly idealized. 
In practice, it is not necessary for individual $i$ to always outperform individual $j$; it suffices if the probability that $i$ performs better than $j$---that is, $v({\mathbf S}\cup \overline{i}\setminus \overline{j}) \ge v({\mathbf S}\cup \overline{j}\setminus \overline{i})$---is significantly high for specific roles, or if $i$ is a fast learner capable of quickly acquiring the necessary skills.
Factors such as education, interest, motivation, and experience all influence this probability.
Skilled job interviewers can often estimate these probabilities effectively.

In large and medium-sized economies, it is impractical to use $\gamma_i[v]$ or $\lambda_i[v]$ to allocate individual welfare or benefits, respectively. The complexity and scale of economic activities make it difficult to estimate these values accurately. Instead, individual payoffs can be determined by market forces, while ensuring that overall employment welfare remains within the desired range of $(1-\tau)v(\mathbf{S})$ and aggregate unemployment benefits are approximately $(\tau-\delta)v(\mathbf{S})$. By making reasonable assumptions about productivity levels and EEO, a more nuanced approach can be devised. Although this method overlooks various economic factors and individual contributions, it allows for a simpler allocation.

In the absence of detailed analysis or prior knowledge about the production function $v$, assuming symmetry among the unemployed (or the employed) offers a reasonable {a priori} basis for distributing unemployment benefits (or employment welfare, respectively).
For instance, the Cobb-Douglas production function presumes symmetry among all employed individuals. 
In many real-world systems---such as in the United States---unemployment insurance benefits are not strictly proportional to individual contributions. Instead, benefits are typically determined by recent earnings history, state-specific formulas and caps, and eligibility criteria. 
These systems operate more as social insurance mechanisms than private insurance models, prioritizing income smoothing and poverty prevention over strict actuarial fairness.

Both group symmetry and EEO embody the principle of the veil of ignorance. 
Under this hypothetical veil, policymakers are unaware of their own social status, wealth, employment condition, productivity, or other personal characteristics.
This perspective compels them to design a fair system that applies equally to all individuals, regardless of their future circumstances (e.g., \cite{Rawls1999}). 
EEO can also be justified by the principle of insufficient reason (also known as the principle of indifference), which assigns equal probabilities to all competing scenarios of $\mathbf{S}$ when there is no basis for favoring one over~another.

Assuming EEO and symmetry among both employed and unemployed individuals, the $\phi$-rate tax $\tau(\omega, \delta) = 1 - \omega + \delta \omega$ effectively eliminates income inequality when income is defined as the sum of employment welfare and unemployment benefits. In this framework, labor costs are treated as reimbursements for the time, effort, learning, commute, and other employment-related expenses incurred by employees. 
Theorem \ref{thm:equality_of_outcome} confirms this equality of outcome. 
Notably, the theorem does not impose any restrictions on the size of the labor force $n$, nor on the specific probability distribution used to model EEO---although the $\phi$-rate is derived from the EEO condition defined in Equation~(\ref{eq:probabilitydensity}).

\begin{theorem}\label{thm:equality_of_outcome}
Assume EEO and symmetry among the employed in $\mathbb{N}$, and also EEO and symmetry among the unemployed in $\mathbb{N}$.
Then, $\tau(\omega, \delta) = 1 - \omega + \delta \omega$ if and only if employment welfare equals unemployment benefits.
\end{theorem}

\noindent \textbf{Proof}. See Appendix I. \quad $\square$

\medskip
The theorem remains valid even when employed and unemployed individuals differ in their productivity or face unequal employment opportunities.
In practice, those who are employed typically enjoy a higher probability of continued employment and greater productivity than their unemployed counterparts. 
For instance, workers in the technology sector who are currently employed often have access to advanced tools and training, which enhances their productivity compared to unemployed individuals who may lack such resources. Moreover, employed individuals generally benefit from stronger professional networks and better access to information about job openings, further increasing their likelihood of remaining employed.

These disparities are the result of market mechanisms, not of any prior assumptions. The observed differences in employment opportunities and productivity reflect market efficiency and contribute to overall economic performance.
For example, companies may prefer to hire candidates with recent work experience, thereby reinforcing the productivity gap between employed and unemployed groups.

Nevertheless, labor mobility enables asymmetric productivity to gradually diminish when employment opportunities are available.
For example, if a manufacturing firm faces a labor shortage, it can often recruit unemployed workers with similar skill sets. After a brief period of training, these new employees typically achieve productivity levels comparable to those of existing staff. Consequently, a substantial portion of the employed workforce can be replaced by unemployed labor without significantly reducing overall productivity.

\section{Empirical Study, Labor Costs and Other Applications}\label{sect:other_apps}

The preceding analysis has focused on a normative policy framework---one that incorporates value judgments and prescriptive perspectives regarding how policy should function, rather than concentrating solely on empirically testable claims.
As a result, actual tax rates may differ from theoretically fair rates.
Nevertheless, this section demonstrates that, when using specific real-world data, these rates are often closely aligned,  although the degree of proximity may vary depending on the dataset.

Furthermore, we present several alternative applications to highlight the broader utility of the proposed approach.
In abstract terms, the earlier sections offer a fair-division solution within the following game-theoretic context: 
players are randomly divided into two groups, and the random payoff to be allocated originates from one of these groups.
This model has wide-ranging applications. 
We examine four such examples to illustrate the practical relevance of the previously derived formulas. 

Additionally, we introduce methods for estimating labor costs and the reserve ratio $\delta$.

\subsection{Empirical Study}

This study utilizes real-world data to compare $v$-based tax rates with fair rates---defined as $\tau=1-\omega+ \delta\omega$---from 2002 to 2016 for the Nordic countries: Denmark, Finland, Iceland, Norway, and Sweden. 
These nations share many similarities in lifestyle, history, religion, and their social and economic models. 
Throughout the study period, they maintained moderate and stable debt-to-GDP ratios.

All data are sourced from OECD (2025), which includes several categories of taxes, such as those imposed on income (both personal and corporate), consumption, and property.
We define the employment-based, or $v$-based, tax as 
$$
TAX_v = \frac{employment\ compensation}{gross\ value\ added} \times income\ tax.
$$
This represents the portion of income tax attributed to employment, whereas corporate income tax primarily applies to  capital owners.
We approximate production $v$ as employment compensation, net of its tax-exempt amount.
Thus, the $v$-based tax rate is simply $TAX_v / v$, which differs from the average of progressive personal income tax rates.

The reserve $\delta$ relates to expenditures on defense, education, environment protection, general public services, public order and safety, social protection, government deficit repayment, recreation, culture and regional development. We collectively denote their sum as $EXP_\delta$, excluding unemployment-related spending. 
However, only a portion of these expenses are funded by employment-based tax.
On a \textit{pro rata} basis, the employment's proportion of tax revenue, net of unemployment spending, is
$$
\frac{TAX_v}{(total\ tax\ revenue)\ - \ (spending\ on\ unemployment)}.
$$
Therefore, we let: 
$$
\delta = \frac{TAX_v}{(total\ tax\ revenue)\ - \ (spending\ on\ unemployment)}
\times \frac{EXP_\delta}{v}.
$$
The values of $\delta$ vary across countries and years.

\begin{figure}[H]
\centering
\includegraphics[width=.8\textwidth,height=2in]{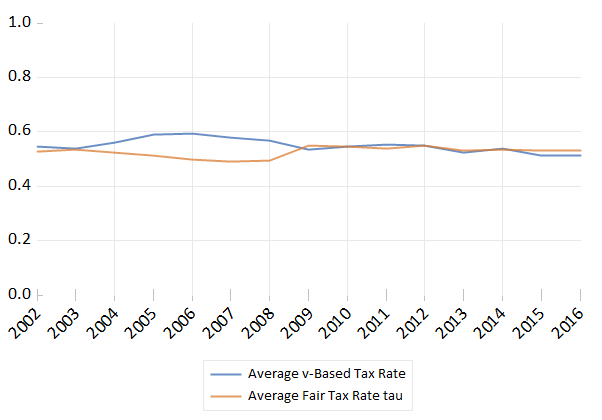}
\caption{Average $v$-based Tax Rate vs Average Fair Tax Rate $\tau$ in Nordic Countries (2002--2016).}
\label{fig:nordic_tax_rates}
\end{figure}

Figure~\ref{fig:nordic_tax_rates} presents both the $v$-based tax rates and the fair rate $\tau$ from 2002 to 2016, averaged across these countries. Although the two rates are not exactly equal, they are sufficiently close and show no systematic bias over most of the years.
 
\vspace{-6pt}

A similar analysis at the individual country level can reveal whether a nation taxes too little or too much, and help explain the reason---either an imbalance in the budget or inequality of employment opportunity.

\subsection{Voting Rights}
In a voting game (e.g., \cite{Shapley1962}), the function $v: 2^\mathbb{N} \to \{0, 1\} $ is a monotonically increasing set function. Let $\mathbf{S}$ denote the random subset of voters who support a proposal. 
The proposal passes if $v(\mathbf{S})=1$; otherwise, it is blocked when $v(\mathbf{S})=0$. 
In this context, however, $v$ does not represent ``production'' or a similar concept. 

According to Hu (2006), $\gamma_i [v]$ denotes the probability that individual $i$ can change an indecisive outcome into a passed one, while $\lambda_i [v]$ represents the probability that $i$ can change an indecisive result into a blocked one. 
The sum $\gamma_i [v]+\lambda_i [v]$ quantifies individual $i$'s political power---that is, their influence over outcome. 

The ratio $\delta$ becomes relevant in specific scenarios. 
For example, suppose $5\%$ of voters initiate a petition to hold a referendum on a proposal.
These petitioners are not part of the random subset $\mathbf{S}$, as their votes are predetermined. 
Collectively, they hold $\delta=0.05$ of the total voting power.

Many voting games are symmetric, particularly those governed by the principle of ``One Person, One Vote.''
In such cases, equality of outcome corresponds to an egalitarian distribution of voting power.

\subsection{Health Insurance}\label{subsect:Insurance}
Health insurance typically involves two types of policyholders: those who are sick and utilize the insurance to cover medical expenses, and those who are healthy and do not make use of it.  
Let $\mathbf{S}$ denote the random set of sick policyholders, $v(\mathbf{S})$ represent the total medical expenses (after deducting copays), and $\tilde \delta v(\mathbf{S})$ be the surcharge paid to the insurance company. 
Define $\delta = - \tilde \delta$. 
The total expenses, excluding copays, amount to $(1 - \delta) v(\mathbf{S})$, which are distributed among all policyholders.

If $\tau=1-\omega+\delta\omega$, and the function $v$ is symmetric across two types of policyholders, then---by the principle of equality of outcome---the cost of purchasing an insurance policy~is: 
$$
\frac{(1-\delta) \mathbb{E} \left [ v(\mathbf{S}) \right ]}{n}.
$$ 
Here, we take the expectation of $v(\mathbf{S})$ because policyholders pay this cost upfront. 
In contrast, unemployment benefits and employment welfare are distributed after the realization of $v(\mathbf{S})$.

In practice, the principle of equality of outcome is applied, and thus the rule \mbox{$\tau=1-\omega+\delta\omega$} holds.
In this scenario, patients also pay predetermined copays. 
Since the same copay applies to all patients regardless of illness severity, it can be set based on the minimum medical cost incurred when any policyholder becomes sick.

\subsection{Dynamic Highway Toll}\label{subsec:highway_toll}
The I-66 highway segment inside the Capital Beltway (I-495) utilizes a dynamic tolling system during rush hours: carpool drivers travel toll-free, while each solo driver pays a dynamic toll, denoted as $\xi (n, s)$, where $n$ is the total number of drivers and $s$ is the number of solo drivers in the segment.
Let $\mathbf{S}$ represent the set of solo drivers.

Let $g(n)$ denote the average cost per driver---ignoring tolls---when the traffic volume is $n$ cars. 
This function increases nonlinearly with $n$. 
A practical estimate for $g(n)$ is the expected driving time (in hours) multiplied by the average hourly wage. 
Thus, the total cost for all drivers is $ng(n)$, and the total cost for carpool drivers---assuming no solo drivers---is $(n-s) g(n-s)$. 
The excess traffic cost caused by solo drivers, after tolls are deducted, is:
$$
v(\mathbf{S}) = ng(n) - (n-s) g(n-s) - s \xi (n, s).
$$ 

Since all drivers are symmetric after paying tolls, the principle of equality of outcome implies that each driver should bear the same cost, $v(\mathbf{S})/n$. 
A carpool driver incurs an additional cost of $g(n) - g(n-s)$ due to solo drivers, so we
equate \mbox{$v(\mathbf{S})/n= g(n) - g(n-s)$}. Solving this yields the toll:
\begin{equation}\label{eq:xi_carpool}
\xi (n, s) = g(n-s).
\end{equation}
An administrative surcharge $\delta$ may be added to this base toll, but $s \xi (n, s)$ can represent the administrative charge .

This toll formula in Equation~(\ref{eq:xi_carpool}) is based on per-vehicle equality of outcome.
We can extend this framework to include all carpool passengers, who experience the same traffic conditions as carpool drivers.
Let $c\ge 1$ be the expected number of passengers per carpool vehicle, excluding the driver.
Then, there are expected $n+c(n-s)$ individuals in the segment---including all drivers and passengers---and $n+c(n-s)-s=(c+1)(n-s)$ individuals are exempt from tolls.
The expected excess traffic cost caused by solo drivers---excluding tolls---becomes:
$$
\mathbb{E}\left[v(\mathbf{S})\right] = [n+c(n-s)]g(n) - (c+1)(n-s) g(n-s) - s \xi (n, s).
$$ 
This expected cost is evenly shared among all individuals.
Solving the equation 
$$
\frac{\mathbb{E}\left[v(\mathbf{S})\right]}{n+c(n-s)} = g(n) - g(n-s)
$$ 
leads to the same toll formula as in Equation~(\ref{eq:xi_carpool}), regardless of the value of $c$.
It represents the average saving if any passenger were to pair with a solo driver---converting the solo driver into a carpool driver while keeping all others unchanged.

\subsection{Feature Selection in Machine Learning and Econometrics}

Feature selection, also known as variable selection, is the process of identifying the most relevant predictors from a large set of candidates to model and predict a dependent variable.
Suppose we have $n$ candidate features $X_1, X_2, \cdots, X_n \in \mathbb{N}$, a dependent variable $Y$, and a performance measurement function $v: 2^\mathbb{N}\to R$.
The subset of relevant features $\mathbf{S} \subseteq \mathbb{N}$ is unobservable, and our objective is to estimate it.

Each relevant feature helps explain some aspect of $Y$.
However, if its explanatory power is already captured by more relevant features in $\mathbf{S}$, it should be excluded from $\mathbf{S}$ to avoid overfitting or collinearity.
Thus, a feature outside $\mathbf{S}$ may still be relevant but not necessary for inclusion.

In the absence of prior information, we assume each candidate feature has an equal prior probability of being selected into $\mathbf{S}$; consequently, the final selection is entirely based on the performance of each feature, subject to sampling errors. 
Since $\mathbf{S}$ is unobservable, we simulate it using the following steps to estimate the marginal performance: 
\begin{enumerate}
\item Sample a value $p$ from the beta distribution with hyperparameters $\theta$ and $\rho$.
\item Simulate a size $s$ from the binomial distribution with parameters $n$ and $p$.
\item Simulate $\mathbf{S}$ from the uniform distribution over all subsets of size $s$.
\end{enumerate}
This simulation process depends on the hyperparameters $\theta$ and $\rho$.

The most relevant feature $X_i$ may either be included in the simulated subset $\mathbf{S}$, contributing a marginal gain, or excluded from $\mathbf{S}$, contributing a marginal loss. 
In either case, it influences the value $v(\mathbf{S})$, and the combined marginal contribution---both gain and loss---represents its explanatory or predictive power for $Y$. 
To assess relevance, we use the sample mean, denoted as $\hat \gamma_i[v]+\hat \lambda_i[v]$, of marginal contributions across many simulated $\mathbf{S}$.
In contrast, an irrelevant feature contributes neither marginal gain nor loss to $v(\mathbf{S})$, aside from random noise.

The dependent variable $Y$ is typically not fully determined by the available features. It may include an unknown constant intercept and a random residual, which together represent the $\delta$ share of $v(\mathbf{S})$ not attributed to any feature in $\mathbb{N}$.
In linear regression, for example, we begin with a preliminary regression using all candidate features. 
The resulting $R^2$ value measures the proportion of $Y$'s variability explained by these features. 
Accordingly, we set $\delta=1-R^2$.
From this, Equations~(\ref{eq:tax_rate_def}) and (\ref{eq:tax_rate_c}) imply the following relationship:
\begin{equation}\label{eq:inentification_theta_rho}
\frac{s (\theta+\rho-1)-n\theta}{\rho+n- s-1} + \frac{s (\theta+\rho-1)-n(\theta-1)}{\theta+ s -1} = 1-\delta.
\end{equation}

We sequentially admit relevant features, beginning with the most significant one.
At each step, we select at most one candidate from the remaining pool---specifically, the feature with the greatest and statistically significant  marginal contribution.
Let $\hat{S}$ denote the set of features already admitted, and let $\Xi = \mathbb{N}\setminus \hat{S}$ represent the remaining candidate features.
To select the next feature, we impose the conditions $s=1$, $n=|\Xi|$, and $\mathbb{E}[p]=\frac{\theta}{\theta+\rho}=\frac{s}{n}$ in Equation~(\ref{eq:inentification_theta_rho}), which uniquely determines the values of $\theta$ and $\rho$. Some algebras yields the identified value of $\theta$ as:
$$
\theta = \frac{(n-2)(n-1+\delta) \pm\sqrt{n-2} \sqrt{\delta^2 (n-2) -2\delta n(n-1) + (n+2)(n-1)^2}}{2(1-\delta)(n-1)}
$$
and the identified $\rho=(n-1)\theta$.

Using the identified values of $\theta$ and $\rho$, we simulate numerous subsets $\mathbf{S} \subseteq \Xi$ to estimate the average marginal contribution $\hat\gamma_i[v]+\hat\lambda_i[v]$ for all candidates in $\Xi$.
The feature with the most significant marginal contribution is admitted into $\hat{S}$ and removed from the candidate set $\Xi$ .
We then update $(n, \theta,\rho)$ accordingly.
This process continues until no remaining candidate exhibits statistical significance in marginal contribution. 

The procedure is outlined in Algorithm~\ref{alg:feature_selection}.
In Step 4, $100$ random subsets $\mathbf{S} \subseteq \Xi$ are used to estimate the average marginal contribution.
For a given random subset $\mathbf{S}$, we add one feature from $\Xi\setminus \mathbf{S}$ or drop one feature from $\mathbf{S}$ to determine the feature's marginal contribution. Thus, there are at most $n+1$ model estimations for each random subset. Each feature accumulates $100$ marginal contributions, resulting in a computational complexity of $100n(n+1) = O(n^2)$. 

\begin{algorithm}
\caption{Feature Selection by a Fair Division Rule}\label{alg:feature_selection}
\begin{algorithmic}[1]
\State Initialize:  $\Xi \gets \mathbb{N}$, \ $\hat{S} \gets \emptyset$
\While{$\Xi \not = \emptyset$}		\Comment{admit at most one feature per loop}
\State Set $n \gets |\Xi|$ and update $(\theta,\rho)$
\State Estimate $\hat \gamma_i[v]+\hat \lambda_i[v]$ in the game $(\Xi, v)$ where $v$ also includes regressors from $\hat{S}$
\State Let $j \gets \argmax_{i \in \Xi} \ \hat\gamma_i [v]+\hat\lambda_i[v]$	\Comment{$j$ has the largest marginal contribution}
\If{$\hat\gamma_j [v]+\hat\lambda_j[v]$ is significantly positive}			
\State Add $X_j$ to $\hat{S}$
\State Update $\Xi \gets \mathbb{N} \setminus \hat{S}$  \Comment{bipartition of $\mathbb{N}$ into $\Xi$ and $\hat{S}$}
\Else
\State Return $\hat{\mathbf{S}}$
\EndIf
\EndWhile
\end{algorithmic}
\end{algorithm}

In this algorithm, the performance function $v(S)$ is defined as the negative log-likelihood when the features in $S$ are used to model the dependent variable.
Consequently, the likelihood ratio test is used to assess the statistical significance of $2(\hat\gamma_i[v]+\hat\lambda_i[v])$, which is considered significant if it exceeds the critical value for a $0.05$ significance level under the chi-square distribution with $2$ degrees of freedom.

Simulation studies demonstrate that Algorithm~\ref{alg:feature_selection} significantly outperforms popular feature selection methods such as stepwise, uni-directional, swapwise, combinatorial, auto-search/GETS, and LASSO---all of which are implemented in econometrics software like EViews.
Table~\ref{tbl:feature_selection} compares the precision of these selection methods in identifying a simple linear model.
In this model, there are six candidate features $X_1, \cdots, X_6$. We first simulate $X_1, \cdots, X_5$ as independent normal random variables, and then apply linear transformations to make $X_1, X_2, X_3$ correlate with each other and $X_4$ correlate with $X_5$.
Finally, we set $X_6$ as a linear function of $X_1$ and $X_2$ plus a noise residual, and the dependent $Y$ is a linear model of $X_1$, $X_2$, and $X_3$. Thus, $X_4$ and $X_5$ are irrelevant, while $X_6$ is relevant but not necessarily included in $\mathbf{S}$.

\begin{table}[H]
\centering
\caption{Numbers of Exact Identification in $1000$ Simulated Samples.}\label{tbl:feature_selection}
\resizebox{.9\textwidth}{!}{
\begin{tabular}{r||c c c c c c|c}\hline
\textbf{Method}&\textbf{Stepwise}&\textbf{Uni-Directional}&\textbf{Swapwise}&\textbf{Combinatorial}&\textbf{Auto-Search/GETS}&\textbf{LASSO}&\textbf{Algorithm~\ref{alg:feature_selection}}\\ \hline
\textbf{Precision}&119&119&0&0&119&91&446\\	\hline
\end{tabular}
}
\end{table}

\subsection{Labor Costs}
Sections~\ref{subsect:Insurance} and \ref{subsec:highway_toll} address minimal medical costs and equality of outcome, respectively, to help calibrate the costs excluded from the function $v({\mathbf S})$. 
An alternative approach considers both the demand and supply sides of the labor market.
For example, the labor cost for any $i \in {\mathbf S}$ can be defined as the minimum wage required for all $j \in \mathbb{N}\setminus {\mathbf S}$, where $j$ is either symmetric to or uniformly outperforms $i$ in terms of $v$.
In other words, someone from $\mathbb{N}\setminus {\mathbf S}$ can replace $i$ without reducing the production value $v$.
This minimum acceptable wage is known as the \textit{reservation wage}, below which $j$ is unwilling to work. 

At this market replacement cost, the subset ${\mathbf S}$ can substitute $i$ with $j$ without sacrificing net profit,
and $j$ is willing to accept the job (e.g., \cite{Horowitz&Mcconnell2003}). 
Using the prevailing market minimal wage implies that raises or promotions become necessary when market conditions change or when employees acquire more relevant skills and experience.

To avoid disincentivizing work, unemployment benefits must be lower than the sum of the reservation wage and employment welfare.
To sustain a high labor participation rate, other incentive compatibility measures may include a minimum wage floor for labor costs and sufficient unemployment benefits to cover the unpaid efforts involved in job searching and skill development, both of which contributes positively to market efficiency.

Under the symmetry and EEO assumptions outlined in Theorem~\ref{thm:equality_of_outcome}, Corollary~\ref{cr:equality_of_outcome} asserts that the average employment welfare should never exceed the average unemployment benefits for any given $\phi$-rate.
Given the previously discussed advantages, the tax rate $\tau=1-\omega+\delta\omega$ is likely to be adopted.
Consequently, when labor costs are considered, employed individuals continue to be better off than unemployed participants in the \mbox{labor force.}

\begin{corollary}\label{cr:equality_of_outcome}
Under the assumptions in Theorem~\ref{thm:equality_of_outcome},
the average employment welfare is no greater than the average unemployment benefits for any $\phi$-rate.
\end{corollary}

\noindent \textbf{Proof}. See Appendix J. \quad $\square$

\medskip
Labor costs also indirectly affect the tax rate $\tau$ when the reserve $\delta v(\mathbf{S})$ is held constant.
As labor costs rise, $v(\mathbf{S})$ decreases. 
If $\delta v(\mathbf{S})$ remains unchanged, then $\delta$ must increase, which in turn raises $\tau=1-\omega+\delta\omega$.
For example, in the context of health insurance, a higher copay reduces the cost of purchasing the policy.
Furthermore, the government could implement a countercyclical fiscal policy by adjusting the spending level $\delta$ in response to changes in $\omega$, such that a higher tax rate corresponds to a higher employment rate, i.e.,
$$
\frac{\mathrm{d} \tau}{\mathrm{d} \omega} 
= \frac{\mathrm{d} (1-\omega+\delta\omega)}{\mathrm{d} \omega}
= -1 + \delta + \omega \frac{\mathrm{d} \delta}{\mathrm{d} \omega} > 0.
$$
This inequality allows $\delta$ to increase with $\omega$. 
Its change rate above $(1-\delta)/\omega$ supports a countercyclical policy.

In practice, calibrating reservation wages for thousands of job titles can be prohibitively challenging or costly.
One solution is to establish a universal tax-exempt threshold based on the minimum reservation wage across all occupations.
Alternatively, the threshold could be determined by multiplying the minimum hourly wage by the average yearly working hours.
Another option is to set the threshold according to the poverty line or the living wage standard.

\section{Policy Implications and Discussion}\label{sect:conclusion}

This paper presents a fair-division framework for allocating unemployment benefits in an economy characterized by a largely unknown, heterogeneous-agent production function and a labor market divided into two random groups.
Fairness is defined as equal employment opportunity (EEO), which is modeled using a beta-binomial probability distribution.
While this distribution facilities the derivation of a key tax rate, many of the paper's findings remain robust regardless of its specific form.

To ensure a balanced taxation budget---free from deficits and surpluses---the dichotomous valuation framework proposed by Hu (2006, 2020) is employed. 
By assigning value to unemployed labor, this approach describes how net production is distributed between employed and unemployed individuals. 

By integrating policy variables with Bayesian hyperparameters, this paper identifies a sustainable tax policy through the minimization of the asymptotic variance of the posterior employment rate.
Alternatively, this policy can be uniquely determined by maximizing the posterior mean of the employment rate, minimizing downside risk, or reducing the posterior mean absolute deviation. 

The proposed tax rule is both practical and transparent.
It incentivizes unemployed individuals to seek employment and encourages those already employed to enhance their productivity.
Additionally, under the assumption of symmetric productivity, the model achieves equality of outcome, thereby contributing to a reduction in income inequality.

This framework is policy-oriented. When designing the tax rate, key policies---such as EEO and balanced budget---are prioritized as objectives or constraints. The derived results, including equality of outcome, have implications that extend to other policy areas.
Table~\ref{tbl:policy_implication} summarizes several related policy implications.

\begin{table}[H]
\centering
\caption{Summary of Policy Implications}\label{tbl:policy_implication}
\resizebox{.9\textwidth}{!}{
\begin{tabular}{r|l}\hline
\textbf{Policy Implication}&\textbf{Description}\\ \hline\hline
Fair, sustainable benefits&Based on EEO, not on equality of productivity or outcome\\ \hline 
Data-driven payroll tax&Depends only on unemployment rate and public reserve ratio\\ \hline 
Balanced budget rule&No borrowing/lending; prevents deficits and surpluses\\ \hline 
Equality of opportunity&Views unemployment as efficiency, not failure\\ \hline 
Robustness and simplicity&Minimal assumptions; relies only on observable data\\ \hline 
Reducing inequality \& poverty&Equality of outcome reduces economic inequality\\ \hline 
Applicability to other domains&Extends to voting, insurance, tolling, machine learning\\ \hline 
Proactive government strategies&Includes job creation, training, matching platforms\\ \hline 
Value added versus labor cost&Tax value added only; labor cost should be tax-exempt\\ \hline 
Transparency and predictability&Predetermined; reduces political influence \& government size\\ \hline
\end{tabular}
}
\end{table}

While the allocation rule provides a foundational perspective, it overlooks several critical aspects of real-world labor markets.
First, it does not account for the dynamic nature of income inequality or the rational response to tax policies. 
Second, although restricting the fungibility of future borrowing can help mitigate the risk of national debt accumulation, it also constrains the government's ability---particularly through monetary policy---to intervene effectively in the economy. 
Nevertheless, moderate economic stimulation during recessions remains feasible by adjusting the reserve ratio. 
Third, the EEO assumption does not reflect recent developments in incomplete-market theory (e.g., \cite{Magill&Quinzii1996}). 
Additionally, employing a multi-criteria objective function may be more appropriate than relying solely on a minimum-variance approach, especially in contexts characterized by high unemployment or a large reserve ratio. 
Lastly, the use of a single tax rule may oversimplify the complexities of real-world taxation systems, which are influenced by numerous additional factors. 

These limitations highlight opportunities for further development of the current framework.
Several potential extensions merit exploration. 
For example, the probability distribution used to model EEO could be redefined in various ways. 
One possibility is to replace the two-parameter Beta distribution with a four-parameter version or a beta-rectangular distribution.
Alternatively, the beta-binomial distribution could be substituted with a Dirichlet-multinomial or a beta-geometric distribution. 
Each alternative would require additional identification restrictions to precisely determine the function $\tau(\omega, \delta)$.

Another approach is to allow the hyperparameters $\theta$ and $\rho$ to vary with economic variables, for example, by employing beta regression. 
Additionally, the production function $v$ could be generalized to a multidimensional form, enabling it to capture not only production but also social justice and other societal dimensions.

There are several alternative methods for determining a unique $\phi$-rate, depending on the specific objectives and methodologies adopted.
Statistically, one option is to estimate the structural model using data collected at monthly, weekly, or even daily intervals. 
Other approaches include minimizing the ex-ante risk of $\omega$ or applying the statistical techniques discussed in Section \ref{subsect:FeasibleSet}. 
The objectives may also emphasize characteristics such as skewness, kurtosis, the Gini coefficient, or the entropy of the posterior rate $p_{n, \omega}$.

A game-theoretic approach offers an additional avenue for analysis, particularly through bargaining over the set of feasible solutions $\Omega_{n,\omega,\delta}$, which is especially valuable when $n$ is small. 
Policymakers may also consider modeling the reserve ratio $\delta$ as a function of $\omega$, or prioritizing marginal gains over marginal losses to more effectively stimulate job growth.

Different priors and identification strategies may yield varying results.
Ideally, the outcomes should be robust---minimally sensitive to the choice of priors and consistent across different identification methods.

In summary, the tax policy proposed in this study is grounded in strong theoretical foundations. 
It is straightforward to implement, promotes both productivity and employment, and performs well across a range of objectives. When applying this framework in real-world contexts, it is essential to consider alternative models of equal opportunity, incorporate diverse evaluation criteria, introduce additional constraints, and account for dynamic elements.

\vspace{.5in}

\noindent \textbf{Acknowledgments:}
The author thanks anonymous referees for insightful suggestions.

\noindent \textbf{Supplemental Material:}
An EViews program is available at \url{https://www.mdpi.com/article/10.3390/g16060066/s1} which produces Table~\ref{tbl:feature_selection}, comparing the accuracy of the feature selection approach in Algorithm~\ref{alg:feature_selection} with those implemented in EViews. Users can extend the linear model with additional regressors and expect similar results.

\noindent \textbf{Abbreviations}:
The following abbreviations are used in this manuscript:

\noindent {\footnotesize
\begin{longtable}[c]{@{}ll}
$\mathbb{N}=\{1,2,\cdots,n\}$&Set of all participants in the labor force, labeled as $1, 2,\cdots,n$\\
$\mathbf{S} \subseteq \mathbb{N}$&Random subset of employed participants\\
$S\subseteq \mathbb{N}$&A realization of $\mathbf{S}$\\
$|T|$ &Cardinality of the subset $T\subseteq \mathbb{N}$\\
$n$ &Cardinality of $\mathbb{N}$, i.e., $|\mathbb{N}|$\\
$t$ &Cardinality of the set $T\subseteq \mathbb{N}$, i.e., $|T|$\\
$z$ &Cardinality of the set $Z\subseteq \mathbb{N}$, i.e., $|Z|$\\
$s$ &Either $|\mathbf{S}|$ or its realization $|S|$\\
$\omega=s/n$&Employment rate, i.e., $1 - $ unemployment rate\\
$\overline{i}$&The singleton set $\{ i \}$\\
$\setminus$&Set subtraction\\
$\cup$&Set union\\
$\beta(\theta,\rho)$&The beta function with parameters $\theta$ and $\rho$\\
EEO&Equal Employment Opportunity, or Equality of Employment Opportunity\\
$p$&Probability for the binomial distribution of $|\mathbf{S}|$, which has a beta distribution\\
$(\theta,\rho)$&Parameters of the beta distribution\\
$\mathbb{P}(\mathbf{S}=T)$&Probability of $\mathbf{S}$ equals $T$\\
$p_{_{n,\omega}}$&Posterior employment rate with a updated beta distribution $(\theta+s,\rho+n-s)$\\
$v:2^\mathbb{N}\to R$&Heterogeneous-agent production function, net of labor cost\\
$\emptyset$&Empty set, where $v(\emptyset) = 0$\\
$\eqdef$&Definition of a math formula\\
$\mathbb{E}[\cdot]$&Expectation with respect to the probability distribution of $\mathbf{S}$, e.g., $\mathbb{E}[v(\mathbf{S})]$\\
$\gamma_i[v]$&Individual $i$'s expected marginal gain in $\mathbf{S}$, see Equation~(\ref{eq:gamma})\\
$\lambda_i[v]$&Individual $i$'s expected marginal loss to $\mathbf{S}$, see Equation~(\ref{eq:lambda})\\
$\delta$&Reserve ratio of the production used for the public\\ 
$\tau(\omega,\delta,n)$&Tax rate satisfying both EEO and balanced budget rule, see Equation~(\ref{eq:tax_rate_def})\\
$\Omega_{\omega,\delta,n}$&Feasible set of $(\theta,\rho, \tau)$ which satisfies both EEO and budget constraints\\
$\tau(\omega,\delta)$&Limit of $\tau(\omega,\delta,n)$ as $n\to \infty$, where $(\theta,\rho,\tau)\in \Omega_{\omega, \delta, n}$\\
$\phi$-rate&Limit of tax rates that satisfy Equations~(\ref{eq:tax_rate_def}) and (\ref{eq:tax_rate_c}) as $n\to \infty$\\
$\Delta, \Delta_1, \cdots, \Delta_9$&Shorthand notations in terms of $\omega$, $\tau$, and $\delta$\\
MAD&Mean Absolute Deviation from the mean\\
$R$&Set of real numbers from $-\infty$ to $\infty$\\
$R^2$&Coefficient of determination, i.e., the fraction of the variation in the \\
     &dependent variable which is explained by the independent variables
\end{longtable}
}

\subsection*{Appendix A. Proof of Theorem~\ref{thm:total_D_value}}\label{appd:A}

In this proof, we utilize the following properties of the beta function:
$$
\left \{
\begin{array}{lcll}
\beta (x-1, y+1) &=& \frac{y}{x-1} \beta (x, y), 	&\forall x>1,\ y>0; \\
\beta (x+1, y-1) &=& \frac{x}{y-1} \beta (x,y), 	&\forall x>0,\ y>1. \\
\end{array}
\right .
$$

First, the expected aggregate marginal gain and loss are given by:
$$
\begin{array}{rcl}
\mathbb{E}\left[ \sum\limits_{i\in \mathbf{S}} \left ( v(\mathbf{S})-v(\mathbf{S}\setminus \overline{i}) \right ) \right ]	
&=&
\sum\limits_{S\subseteq \mathbb{N}} \mathbb{P}(\mathbf{S}=S) \sum\limits_{i\in S} \left [ v(S)-v(S\setminus \overline{i}) \right ] \\
&=&
\sum\limits_{i \in \mathbb{N}} \ \sum\limits_{S\subseteq \mathbb{N}: i \in S} \mathbb{P}(\mathbf{S}=S)  \left[ v(S)-v(S\setminus \overline{i}) \right]
=
\sum\limits_{i \in \mathbb{N}} \gamma_i[v], \\
	
\mathbb{E}\left[ \sum\limits_{i \in \mathbb{N}\setminus \mathbf{S}} \left ( v(\mathbf{S}\cup \overline{i})-v(\mathbf{S}) \right ) \right ]
&=&
\sum\limits_{S\subseteq \mathbb{N}} \mathbb{P}(\mathbf{S}=S) \sum\limits_{i \in \mathbb{N}\setminus S} \left[ v(S\cup \overline{i})-v(S) \right] \\
&=&
\sum\limits_{i \in \mathbb{N}} \ \sum\limits_{S \subseteq \mathbb{N}: i \not \in S} \mathbb{P}(\mathbf{S}=S)  \left [ v(S\cup \overline{i})-v(S) \right ] 
=
\sum\limits_{i \in \mathbb{N}} \lambda_i[v].	
\end{array}
$$

Applying Equations~(\ref{eq:probabilitydensity})--(\ref{eq:lambda}), 
the expected marginal gain and loss can be rewritten \mbox{as follows:}
\begin{equation} \label{eq:4expectations}\tag{A1}
\begin{array}{rcl}
\gamma_i [v]
&=& 
\sum\limits_{S\subseteq \mathbb{N}: S\ni i} \frac{\beta(\theta+s,\rho+n-s)}{\beta(\theta,\rho)}[v(S)-v(S\setminus\overline{i})] \\
&\stackrel{Z=S\setminus\overline{i}}=& 
\sum\limits_{S\subseteq \mathbb{N}: S\ni i} \frac{\beta(\theta+s,\rho+n-s)}{\beta(\theta,\rho)} v(S)
- \sum\limits_{Z\subseteq \mathbb{N} \setminus \overline{i}} \frac{\beta(\theta+z+1, \rho+n-z-1)}{\beta(\theta,\rho)} v(Z), \\
		
\lambda_i [v]
&=&
\sum\limits_{Z\subseteq \mathbb{N}\setminus \overline{i}} \frac{\beta(\theta+z,\rho+n-z)}{\beta(\theta,\rho)}[v(Z\cup \overline{i}) -v(Z)] \\
&\stackrel{S=Z\cup \overline{i}}{=}&
\sum\limits_{S\subseteq \mathbb{N}: S\ni i}
\frac{\beta(\theta+s-1,\rho+n-s+1)}{\beta(\theta,\rho)} v(S) 
-
\sum\limits_{Z\subseteq \mathbb{N}\setminus \overline{i}} \frac{\beta(\theta+z,\rho+n-z)}{\beta(\theta,\rho)} v(Z). 
\end{array}
\end{equation}

From Equation~(\ref{eq:4expectations}), the aggregate value of employed labor is given by:
$$
\begin{array}{rcl}
\sum\limits_{i \in \mathbb{N}} \gamma_i [v]
&=&
\sum\limits_{i\in \mathbb{N}} \ \sum\limits_{S\subseteq \mathbb{N}: S\ni i} \frac{\beta(\theta+s, \rho+n-s)}{\beta(\theta,\rho)}v(S)
- \sum\limits_{i\in \mathbb{N}} \ \sum\limits_{Z\subseteq \mathbb{N} \setminus \overline{i}}  \frac{\beta(\theta+z+1, \rho+n-z-1)}{\beta(\theta,\rho)} v(Z) \\
	
&\stackrel{S=Z}=& 
\sum\limits_{S\subseteq \mathbb{N}: S\not = \emptyset} \ \sum\limits_{i\in S} \frac{\beta(\theta+s, \rho+n-s)}{\beta(\theta,\rho)}v(S)
- \sum\limits_{S\subseteq \mathbb{N}: S\not = \mathbb{N}} \ \sum\limits_{i \in \mathbb{N} \setminus S} \frac{\beta(\theta+s+1, \rho+n-s-1)}{\beta(\theta,\rho)} v(S)\\
	
&=&
\sum\limits_{S\subseteq \mathbb{N}: S\not = \emptyset}\frac{s \beta(\theta+s, \rho+n-s)}{\beta(\theta,\rho)}v(S)
- \sum\limits_{S\subseteq \mathbb{N}: S\not = \mathbb{N}} \frac{(n-s) \beta(\theta+s+1, \rho+n-s-1)}{\beta(\theta,\rho)} v(S) \\
	
&=& 
\frac{n\beta(\theta+n,\rho)}{\beta (\theta, \rho)}v(\mathbb{N})
- \frac{n\beta(\theta+1,\rho+n-1)}{\beta (\theta, \rho)}v(\emptyset) \\ 
&&
+ \sum\limits_{S\subseteq \mathbb{N}: S\not = \emptyset, S\not = \mathbb{N}}
\frac{s \beta(\theta+s, \rho+n-s) - (n-s) \beta(\theta+s+1, \rho+n-s-1)}{\beta(\theta,\rho)}v(S) \\
	
&=& 
\frac{n\beta(\theta+n,\rho)}{\beta(\theta,\rho)}v(\mathbb{N}) 
- \frac{n\theta}{\rho+n-1}\frac{\beta(\theta,\rho+n)}{\beta (\theta, \rho)}v(\emptyset) \\
&&
+ \sum\limits_{S\subseteq \mathbb{N}: S\not = \mathbb{N}, S\not = \emptyset}\left[s - \frac{(n-s) (\theta+s)}{\rho+n-s-1} \right] \frac{\beta(\theta+s, \rho+n-s)}{\beta(\theta,\rho)}v(S) \\
	
&=& 
\sum\limits_{S\subseteq \mathbb{N}} \frac{s(\theta+\rho-1)-n\theta}{\rho+n-s-1}  \
\frac{\beta(\theta+s, \rho+n-s)}{\beta(\theta,\rho)} \ v(S).
	
\end{array}
$$

Similarly, the aggregate value of unemployed labor is given by:
$$
\begin{array}{rcl}
\sum\limits_{i\in \mathbb{N}} \lambda_i [v]
&=&
\sum\limits_{i\in \mathbb{N}} \ \sum\limits_{S\subseteq \mathbb{N}: S\ni i} \frac{\beta(\theta+s-1, \rho+n-s+1)}{\beta(\theta,\rho)}v(S)
- 
\sum\limits_{i\in \mathbb{N}} \ \sum\limits_{Z \subseteq \mathbb{N} \setminus \overline{i}} \frac{\beta(\theta+z, \rho+n-z)}{\beta(\theta,\rho)} v(Z) \\
	
&\stackrel{S=Z}=& 
\sum\limits_{S\subseteq \mathbb{N}: S\not = \emptyset} \ \sum\limits_{i\in S} \frac{\beta(\theta+s-1, \rho+n-s+1)}{\beta(\theta,\rho)}v(S)
-  
\sum\limits_{S\subseteq \mathbb{N}: S \not =\mathbb{N}} \ \sum\limits_{i\in \mathbb{N} \setminus S} \frac{\beta(\theta+s, \rho+n-s)}{\beta(\theta,\rho)} v(S) \\
	
&=& 
\sum\limits_{S\subseteq \mathbb{N}: S\not = \emptyset} \frac{s\beta(\theta+s-1, \rho+n-s+1)}{\beta(\theta,\rho)}v(S)
-  
\sum\limits_{S\subseteq \mathbb{N}: S\not =\mathbb{N}}\frac{(n-s)\beta(\theta+s, \rho+n-s)}{\beta(\theta,\rho)} v(S) \\
	
&=& 
\frac{n\beta(\theta+n-1, \rho+1)}{\beta(\theta,\rho)}v(\mathbb{N})
- 
\frac{n\beta(\theta, \rho+n)}{\beta(\theta,\rho)} v(\emptyset) \\
	
&&
+\sum\limits_{S\subseteq \mathbb{N}: S\not = \emptyset, S \not =\mathbb{N}}
\frac{s\beta(\theta+s-1, \rho+n-s+1)-(n-s)\beta(\theta+s, \rho+n-s)}{\beta(\theta,\rho)}v(S) \\
	
&=& 
\frac{n\rho}{\theta+n-1}\frac{\beta(\theta+n, \rho)}{\beta(\theta,\rho)}v(\mathbb{N})
- \frac{n\beta(\theta, \rho+n)}{\beta(\theta,\rho)} v(\emptyset) \\
	
&& 
+\sum\limits_{S\subseteq \mathbb{N}: S\not = \emptyset, S\not = \mathbb{N}} \left[ \frac{s(\rho+n-s)}{\theta+s-1} - (n-s) \right]
\frac{\beta(\theta+s, \rho+n-s)}{\beta(\theta,\rho)} v(S)\\
	
&=& 
\sum\limits_{S\subseteq \mathbb{N}} \frac{s(\theta+\rho-1)-n(\theta-1)}{\theta+s-1} \
\frac{\beta(\theta+s, \rho+n-s)}{\beta(\theta,\rho)} \ v(S).
\end{array}
$$

\subsection*{Appendix B. Proof of Lemma~\ref{lm:solvethetarho_delta}}\label{sect:proof_lemma}\label{appd:B}

For simplicity, we introduce the following shorthand notations:
$$
\begin{array}{lcl}
\Delta_5&\equiv&- \delta\omega + \delta\tau - 2 \delta  + \omega - \tau^2 + 4\tau -2 \equiv \Delta_2+\Delta_4, \\
\Delta_6&\equiv&-\omega \delta + \omega\tau+\tau-1	\equiv \Delta_2 + \omega \Delta_3, \\
\Delta_7&\equiv&-\delta - \omega\tau + 2\tau + \omega - 1 \equiv \Delta_4 + (1-\omega) \Delta_3, \\
\Delta_8&\equiv&-\delta\omega - \delta +  \omega + 3\tau -2 \equiv \Delta_3+\Delta_5, \\
\Delta_9&\equiv& -2\delta\omega  -\delta +2\omega + 4\tau -3 \equiv \Delta + \Delta_8.
\end{array}
$$
All these expressions are bounded since $0\le \omega, \tau, \delta \le 1$. 

Letting $s=n\omega$, we rewrite Equations~(\ref{eq:tax_rate_def}) and (\ref{eq:tax_rate_c}) as a linear system in the unknowns $(\theta, \rho)$:
$$
\left \{
\begin{array}{rcl}
(1-\tau)(\rho+n-s-1) 
&=&
s(\theta+\rho-1)-n\theta, \\
	
(\tau-\delta)(\theta+s-1)
&=&
s(\theta+\rho-1)-n(\theta-1).
\end{array}
\right .
$$
This system has a unique symbolic solution for $(\theta, \rho)$.
We now verify that Equation~(\ref{eq:theta_rho_in_tau}) satisfies this system.

Assuming Equation~(\ref{eq:theta_rho_in_tau}), we derive the following identities (also used in other proofs):
$$
\begin{array}{rcl}
	
\theta+s
&=&
\frac{n^2 \omega \Delta_1 + n\Delta_2 +\Delta_3 }{n\Delta + \Delta_3}
+ \frac{n\omega (n\Delta + \Delta_3)}{n\Delta + \Delta_3} 
=
\frac{n^2 \omega + n\Delta_6 + \Delta_3}{n\Delta + \Delta_3}, \\
	
\theta+s-1
&=&
\frac{n^2 \omega + n\Delta_6 + \Delta_3}{n\Delta + \Delta_3}
- \frac{n\Delta + \Delta_3}{n\Delta + \Delta_3} 
=
\frac{n^2 \omega + n (\Delta_6-\Delta) } 
{n\Delta + \Delta_3}
=
\frac{n^2 \omega + n \omega (\tau - 1)} 
{n\Delta + \Delta_3}, \\
	
\theta+\rho
&=&
\frac{n^2 \omega \Delta_1 + n\Delta_2 +\Delta_3 }{n\Delta + \Delta_3} 
+
\frac{n^2 (1-\omega) \Delta_1 + n\Delta_4 + \Delta_3}{n\Delta + \Delta_3} 
=
\frac{n^2 \Delta_1	+ n \Delta_5	+ 2\Delta_3}{n\Delta + \Delta_3}, \\
	
\theta+\rho - 1
&=&
\frac{n^2 \Delta_1	+ n \Delta_5	+ 2\Delta_3}{n\Delta + \Delta_3}
- \frac{n\Delta + \Delta_3}{n\Delta + \Delta_3} 
=
\frac{	n^2 \Delta_1 + n ( \delta\tau - 2 \delta   - \tau^2 + 3\tau - 1) + \Delta_3} 
{n\Delta + \Delta_3}, \\
	
\theta+\rho + n
&=&
\frac{n^2 \Delta_1	+ n \Delta_5	+ 2\Delta_3}{n\Delta + \Delta_3}
+ \frac{n(n\Delta + \Delta_3)}{n\Delta + \Delta_3} 
=
\frac{n^2 + n \Delta_8 + 2\Delta_3}{n\Delta + \Delta_3}, \\
	
\theta+\rho + n + 1
&=&
\frac{n^2 + n \Delta_8 + 2\Delta_3}{n\Delta + \Delta_3}
+\frac{n\Delta + \Delta_3}{n\Delta + \Delta_3} 
=
\frac{n^2 + n \Delta_9 + 3\Delta_3}{n\Delta + \Delta_3}, \\
	
\theta+\rho + n - 1
&=&
\frac{n^2 + n \Delta_8 + 2\Delta_3}{n\Delta + \Delta_3}
- \frac{n\Delta + \Delta_3}{n\Delta + \Delta_3} 
=
\frac{n^2 + n (\Delta_8 - \Delta) + \Delta_3}{n\Delta + \Delta_3}, \\
	
\theta+\rho + n - 2
&=&
\frac{n^2 + n \Delta_8 + 2\Delta_3}{n\Delta + \Delta_3}
- \frac{2(n\Delta + \Delta_3)}{n\Delta + \Delta_3}
=
\frac{n^2 + n (\Delta_8-2\Delta)}{n\Delta + \Delta_3}, \\
	
\rho+n-s
&=&
\frac{n^2 (1-\omega) \Delta_1 + n\Delta_4 + \Delta_3}{n\Delta + \Delta_3}
+ \frac{n(1-\omega)(n\Delta + \Delta_3)}{n\Delta + \Delta_3} 
=
\frac{n^2 (1-\omega) + n \Delta_7 + \Delta_3}{n\Delta + \Delta_3}, \\
	
\rho+n-s - 1
&=&
\frac{n^2 (1-\omega) + n \Delta_7 + \Delta_3}{n\Delta + \Delta_3}
-\frac{n\Delta + \Delta_3}{n\Delta + \Delta_3} 
=
\frac{n^2 (1-\omega)+ n (1-\omega)(\tau-\delta)}{n\Delta + \Delta_3}.
	
\end{array}
$$

we now compute the following expressions:
$$
\resizebox{.9\textwidth}{!}{$
\begin{array}{rcl}
s(\theta+\rho - 1)-n\theta
&=&
\frac{n\omega [n^2 \Delta_1 + n ( \delta\tau - 2 \delta   - \tau^2 + 3\tau - 1) + \Delta_3]} 
{n\Delta + \Delta_3} 
-
\frac{n(n^2 \omega \Delta_1 + n\Delta_2 +\Delta_3)}{n\Delta + \Delta_3}\\
&=&
\frac{n^2[\omega(\delta\tau - 2 \delta   - \tau^2 + 3\tau - 1) -\Delta_2] + n(\omega-1)\Delta_3} 
{n\Delta + \Delta_3} 
=
\frac{n^2(1-\omega)(1-\tau)	+ n(\omega-1)\Delta_3} 
{n\Delta + \Delta_3}, \\
		
s(\theta+\rho - 1)-n(\theta-1)
&=&
\left [s(\theta+\rho - 1)-n\theta \right ] + n \\
	
&=&
\frac{n^2(1-\omega)(1-\tau)	+ n(\omega-1)\Delta_3} 
{n\Delta + \Delta_3}
+
\frac{n(n\Delta + \Delta_3)}{n\Delta + \Delta_3} 
=
\frac{n^2 \omega(\tau-\delta) + n \omega \Delta_3} 
{n\Delta + \Delta_3}. \\
\end{array}
$}
$$

Thus, we verify the identities:
$$
\left \{
\begin{array}{rclcl}
	
\frac{s(\theta+\rho - 1)-n\theta}{\rho+n-s - 1}
&=&
\frac{n^2(1-\omega)(1-\tau)+ n(\omega-1)\Delta_3}{n^2 (1-\omega)+ n (1-\omega)(\tau-\delta)}
&=&
1-\tau,\\
	
\frac{s(\theta+\rho - 1)-n(\theta-1)}{\theta+s-1}
&=& 
\frac{n^2 \omega(\tau-\delta) + n \omega \Delta_3}{n^2 \omega + n \omega (\tau - 1)}
&=&
\tau - \delta. 
\end{array}
\right .
$$

\subsection*{Appendix C. Proof of Theorem~\ref{thm:limit_distribution}}\label{appd:C}
For any integer $z \ge 0$, by the proof of Lemma \ref{lm:solvethetarho_delta}, as $n\to\infty$, we have:
$$
\frac{\theta+s+z}{\theta+\rho+n+z}
=
\frac{\frac{n^2\omega+n\Delta_6+\Delta_3}{n\Delta+\Delta_3}
	+\frac{z(n\Delta+\Delta_3)}{n\Delta+\Delta_3}}
{\frac{n^2+n\Delta_8+2\Delta_3}{n\Delta+\Delta_3}+
	\frac{z(n\Delta+\Delta_3)}{n\Delta+\Delta_3}} 
=
\frac{n^2\omega+n(\Delta_6+z\Delta)+(1+z)\Delta_3}
{n^2+n(\Delta_8+z\Delta)+(2+z)\Delta_3}
\ \to \
\omega.
$$

From Section~\ref{subsect:EEO}, $p_{_{n,\omega}}$ has a Beta distribution with parameters $(\theta+s,\rho+n-s)$.
Thus, its characteristic function is (e.g., \cite{Johnson&Kotz&Balakrishnan1995}, Chapter 21):
$$
\mathbb{E} \left [e^{i \eta p_{_{n,\omega}}} \right ] 
= 1 + \sum\limits_{k=1}^\infty \frac{(i \eta)^k}{k!}
\prod\limits_{z=0}^{k-1} \frac{\theta+s+z}{\theta+\rho+n+z}
$$
where $i$ is the imaginary unit, i.e., $i^2=-1$. 

Taking the limit as $n\to\infty$, we observe:
$$
\lim\limits_{n\to \infty} \mathbb{E}[e^{i \eta p_{_{n,\omega}}}] 
=
1 + \sum\limits_{k=1}^\infty \frac{(i \eta)^k}{k!} \lim\limits_{n\to \infty}\prod\limits_{z=0}^{k-1} 
\frac{\theta+s+z}{\theta+\rho+n+z}
=
1 + \sum\limits_{k=1}^\infty \frac{(i \eta \omega)^k}{k!} 
=
\exp (i \eta \omega).
$$
Therefore, the random variable $p_{_{n,\omega}}$ converges in distribution to a point mass at $\omega$ as $n\to\infty$.

\subsection*{Appendix D. Proof of Theorem~\ref{thm:zero_variance}}\label{appd:D}
To analyze the asymptotic relationship between the functions $f(n)$ and $g(n)$,
we use the following standard notations:
\begin{itemize}
\medskip
\item $f(n)=O \left (g(n) \right )$ if $\limsup\limits_{n\to \infty} \left |\frac{f(n)}{g(n)} \right | < \infty$;
\medskip
\item $f(n) \approx g(n)$ if $\lim\limits_{n\to\infty} \frac{f(n)}{g(n)} =1$.
\medskip
\end{itemize}

Since $p_{_{n,\omega}}$ follows a Beta distribution with parameters $(\theta+s, \rho+n-s)$, its variance is
given by (e.g., \cite{Gupta&Nadarajah2004}, p. 35):
$$
\mathrm{VAR}(p_{_{n,\omega}}) = \frac{(\theta+s)(\rho+n-s)}{(\theta+\rho+n)^2(\theta+\rho+n+1)}.
$$ 
From the proof of Lemma \ref{lm:solvethetarho_delta}, the variance $\mathrm{VAR}(\sqrt{n} p_{_{n,\omega}})$ is:
$$ 
\begin{array}{rcl}
&&
\frac{n \frac{n^2\omega+n\Delta_6+\Delta_3}{n\Delta+\Delta_3}
\frac{n^2(1-\omega)+n\Delta_7+\Delta_3}{n\Delta+\Delta_3}}
{ \left (\frac{n^2+n\Delta_8+2\Delta_3}{n\Delta+\Delta_3}\right )^2
\frac{n^2+n\Delta_9+3\Delta_3}{n\Delta+\Delta_3}} 
=
\frac{n(n\Delta+\Delta_3) 
(n^2\omega+n\Delta_6+\Delta_3)
\left [n^2(1-\omega)+n\Delta_7+\Delta_3 \right ]}
{\left (n^2+n\Delta_8+2\Delta_3 \right )^2
\left (n^2+n\Delta_9+3\Delta_3 \right )} \\
	
&=&
\frac{ \left (\Delta+\frac{\Delta_3}{n}\right )
\left [ \omega \left (1+\frac{\Delta_6}{n\omega}+\frac{\Delta_3}{n^2\omega} \right ) \right ]
\left [ (1-\omega) \left (1+\frac{\Delta_7}{n(1-\omega)}+\frac{\Delta_3}{n^2(1-\omega)} \right ) \right ]}
{   \left (1+ \frac{\Delta_8}{n}+\frac{2\Delta_3}{n^2}\right )^2
\left (1+ \frac{\Delta_9}{n}+ \frac{3\Delta_3}{n^2} \right )} \\
	
&=&
\omega (1-\omega) \left (\Delta+\frac{\Delta_3}{n}\right ) \left [ 1+ \frac{\Delta_6}{n\omega} + \frac{\Delta_7}{n(1-\omega)} - \frac{2\Delta_8}{n}-\frac{\Delta_9}{n}
+ O\left (\frac{1}{n^2} \right ) \right ] \\ 
	
&=&
\omega (1-\omega)\Delta 
+
\frac{\omega (1-\omega)}{n}
\left [ \Delta_3
+ \Delta ( \frac{ \Delta_6 }{\omega} + \frac{\Delta_7}{1-\omega}
- 2\Delta_8 - \Delta_9) \right  ] 
+ O\left (\frac{1}{n^2} \right ) 
\to
\omega (1-\omega) \Delta.
\end{array}
$$
To minimize this asymptotic variance $,\lim\limits_{n\to \infty}\mathrm{VAR}(\sqrt{n} p_{_{n,\omega}}) = \omega (1-\omega)\Delta$,
while ensuring non-negativity, we must set $\Delta= 0$. Therefore, the value of $\tau$ that minimizes the asymptotic variance is:
$$
\argmin_{\tau} \lim\limits_{n\to \infty}
\mathrm{VAR} \left (\sqrt{n} p_{_{n,\omega}} \right ) = 1-\omega+\delta\omega.
$$

\subsection*{Appendix E. Proof of Theorem~\ref{thm:semivariance}}\label{appd:E}
Let $\mu_n = \frac{\theta+s}{\theta+\rho+n}$ be the mean of $p_{_{n,\omega}}$, and let 
$$
f(x) = \frac{x^{\theta+s-1}(1-x)^{\rho+n-s-1}}{\beta ( \theta+s, \rho+n-s)}
$$ 
be its probability density function, defined on $0\le x\le1$.
Let $\sigma_n^2$ be the variance of $p_{_{n,\omega}}$, and define $\alpha_n=\sqrt{\frac{\omega(1-\omega)\Delta}{n}}>0$.

The lower semivariance of $p_{_{n,\omega}}$ is defined as:
$$
\underline{\sigma}^2_n \eqdef \int_0^{\mu_n} f(x) (x-\mu_n)^2 \mathrm{d} x
$$
and $n \underline{\sigma}^2_n$ is the lower semivariance of $\sqrt{n} p_{_{n,\omega}}$.
We apply this in the Chebychev inequality:
\begin{equation}\label{eq:lower1se}\tag{A2}
\begin{array}{r c l}
\mathbb{P} \left(p_{_{n,\omega}} \le \mu_n - \alpha_n \right) 
&=&
\int\limits_0^{\mu_n- \alpha_n} \ f(x) \mathrm{d} x  
\le
\int\limits_0^{\mu_n- \alpha_n} \ f(x) \left( \frac{x - \mu_n}{\alpha_n} \right)^2 \mathrm{d} x \\
&\le&
\frac{1}{\alpha_n^2}\int\limits_0^{\mu_n} f(x)  \left(x - \mu_n\right)^2 \mathrm{d} x 
=
\frac{\underline{\sigma}^2_n}{\alpha_n^2} 
=
\frac{n\underline{\sigma}^2_n}{\omega(1-\omega)\Delta}.
\end{array}
\end{equation}

Let $\kappa_n = \frac{\theta+s-1}{\theta+\rho+n-2}$ and $\varepsilon_n$ denote the mode and median of  $p_{_{n,\omega}}$, respectively.
Since the median lies between the mean and the mode, and using identities from Appendix~\ref{sect:proof_lemma}, \mbox{we have:}
$$
\begin{array}{rcl}
|\mu_n - \varepsilon_n| 
&\le&
|\mu_n - \kappa_n | 
=
\left | \frac{\theta+s}{\theta+\rho+n} - \frac{\theta+s-1}{\theta+\rho+n-2} \right |  
=
\left | \frac{n+\rho-\theta-2s}{(\theta+\rho+n)(\theta+\rho+n-2)} \right | \\
	
&\le& 
\left | \frac{\theta+s}{(\theta+\rho+n)(\theta+\rho+n-2)} \right | 
+
\left | \frac{n+\rho-s}{(\theta+\rho+n)(\theta+\rho+n-2)} \right | 
\ =\ 
O(\frac{1}{n}).
\end{array}
$$

To estimate a lower bound, we use Stirling's approximation for the Gamma function $\Gamma(\cdot)$ where $\Gamma (z+1)\approx \sqrt{2\pi z}\ z^z \ \exp (-z)$. Then, the probability that $p_{_{n,\omega}} \le \mu_n - \alpha_n$ is:

$$
\begin{array}{rcl}
&&
\mathbb{P} \left(p_{_{n,\omega}} \le \mu_n - \alpha_n \right)
= 
\mathbb{P} \left (p_{_{n,\omega}} \le \varepsilon_n \right )
- \int \limits_{\mu_n - \alpha_n}^{\varepsilon_n} f(x) \mathrm{d} x \\
&\ge&
\frac{1}{2}
-
\left  | \varepsilon_n-\left(\mu_n- \alpha_n \right) \right | f(\kappa_n) 
=
\frac{1}{2}
-
\left[ \alpha_n + O(\frac{1}{n}) \right]
\frac{\left (\frac{\theta+s-1}{\theta+\rho+n-2}\right )^{\theta+s-1} 
\left (1-\frac{\theta+s-1}{\theta+\rho+n-2}\right )^{\rho+n-s-1}}{\frac{\Gamma (\theta+s) \Gamma (\rho+n-s)}{\Gamma (\theta+\rho+n)} }\\
&=&
\frac{1}{2}
-
\left[ \alpha_n + O(\frac{1}{n}) \right]
\frac{\frac{(\theta+s-1)^{\theta+s-1}}{\Gamma (\theta+s)}
\frac{(\rho+n-s-1)^{\rho+n-s-1}}{\Gamma (\rho+n-s)}}{\frac{(\theta+\rho+n-2)^{\theta+\rho+n-2}}{(\theta+\rho+n-1) \Gamma (\theta+\rho+n-1)}}\\
&\approx&
\frac{1}{2}
-
\left[ \alpha_n + O(\frac{1}{n}) \right]
\frac{\frac{\exp (\theta+s-1)}{\sqrt{2\pi(\theta+s-1)}}
\frac{\exp (\rho+n-s-1)}{\sqrt{2\pi (\rho+n-s-1)}}}
{\frac{\exp (\theta+\rho+n-2)}{(\theta+\rho+n-1) \sqrt{2\pi (\theta+\rho+n-2)}}} 
=
\frac{1}{2}
-
\left[ \alpha_n + O(\frac{1}{n}) \right]
\frac{(\theta+\rho+n-1) \sqrt{\theta+\rho+n-2}}{\sqrt{2\pi(\theta+s-1)(\rho+n-s-1)}}\\
&\approx&
\frac{1}{2}
-
\left[\sqrt{\frac{\omega(1-\omega)\Delta}{n}}+ + O \left(\frac{1}{n}\right)\right]
\frac{\frac{n}{\Delta} \sqrt{\frac{n}{\Delta}}}
{\sqrt{2\pi \frac{n\omega}{\Delta} \frac{n(1-\omega)}{\Delta}}} 
\ =\
\frac{1}{2}
-
\frac{1}{\sqrt{2\pi}} + O \left (\frac{1}{\sqrt{n}} \right ).
\end{array}
$$

Finally, rewriting Equation~(\ref{eq:lower1se}), we obtain:
$$
\left (\frac{1}{2}
-
\frac{1}{\sqrt{2\pi}} \right )
\omega(1-\omega)\Delta + O \left (\frac{1}{\sqrt{n}} \right ) 
\le 
n \underline{\sigma}^2_n
\le
n \sigma_n^2
=
\omega(1-\omega)\Delta + O \left (\frac{1}{n} \right ).
$$
Taking the limit as $n\to\infty$, we get:
$$
\left (\frac{1}{2}
-
\frac{1}{\sqrt{2\pi}} \right )
\omega(1-\omega)\Delta 
\le 
\liminf\limits_{n\to \infty} n\underline{\sigma}_n^2
\le
\limsup\limits_{n\to \infty} n\underline{\sigma}_n^2
\le
\omega(1-\omega)\Delta.
$$
Therefore, setting $\Delta=0$ minimizes the limiting lower semivariance of $\sqrt{n} p_{_{n,\omega}}$.

A similar argument applies to the upper semivariance of $\sqrt{n} p_{_{n,\omega}}$.

\subsection*{Appendix F. Proof of Theorem~\ref{thm:max_posterior_mean}}\label{appd:F}
Note that:
$$
\Delta_6-\omega\Delta_8 = (1-\tau)(2\omega-1)+\omega^2 (\delta-1).
$$
From the proof of Lemma \ref{lm:solvethetarho_delta}, we have:
$$
\begin{array}{rcl}
&&\mathbb{E} \left[p_{_{n,\omega}} \bigm\vert (\theta,\rho, \tau)\in \Omega_{\omega,\delta,n} \right]
=
\frac{\theta+s}{\theta+\rho+n}
=
\frac{n^2\omega + n \Delta_6 + \Delta_3}{n^2 + n\Delta_8 + 2\Delta_3} \\
&=&
\omega + \frac{n (\Delta_6-\omega\Delta_8) + (1-2\omega) \Delta_3}{n^2 + n\Delta_8 + 2\Delta_3}
=
\omega + \frac{(1-\tau)(2\omega-1)+\omega^2 (\delta-1)}{n} + O(\frac{1}{n^2}).
\end{array}
$$
In this approximation, the mean decreases with increasing $\tau$ when $n$ is large and $\omega>0.5$.
To maximize the mean, we minimize $\tau \in (1-\omega+\delta\omega,1)$ such that $\theta>0$ and $\rho>0$. 

It is straightforward to verify that $\theta>0$ and $\rho>0$ when: 
$$
\tau = 1-\omega+\delta\omega + \frac{2\omega (1-\omega)(1-\delta)^2}{n}
$$ 
and $n$ is sufficiently large. Therefore,
$$
1-\omega+\delta\omega
<
\argmax_{\tau} \ \mathbb{E}\left[p_{_{n,\omega}} \bigm\vert (\theta,\rho, \tau)\in \Omega_{\omega,\delta,n} \right] 
\le
1-\omega+\delta\omega + \frac{2\omega (1-\omega)(1-\delta)^2}{n}.
$$
Taking the limit as $n\to \infty$, we obtain:
$$
\lim\limits_{n\to\infty} \argmax_{\tau} \ \mathbb{E}\left[p_{_{n,\omega}} \bigm\vert (\theta,\rho, \tau)\in \Omega_{\omega,\delta,n} \right]
=
1 -\omega +\delta\omega.
$$

Alternatively, we can reconfirm this result by analyzing the derivative of $\frac{n^2\omega + n \Delta_6 + \Delta_3}{n^2 + n\Delta_8 + 2\Delta_3}$ with respect to $\tau$:
$$
\resizebox{.85\textwidth}{!}{$
\frac{\partial \mathbb{E} \left[p_{_{n,\omega}} \bigm\vert (\theta,\rho, \tau)\in \Omega_{\omega,\delta,n} \right]}{\partial \tau} 
= 
\frac{\left(n\frac{\partial \Delta_6}{\partial \tau}\right)(n^2) - (n^2\omega)\left(n\frac{\partial \Delta_8}{\partial \tau}\right)+O(n^2)}{\left(n^2 + n\Delta_8 + 2\Delta_3\right)^2}
=
\frac{1-2\omega}{n}+O\left(\frac{1}{n^2}\right).
$}
$$
This derivative is negative for $\omega>0.5$ and large $n$, confirming that increasing $\tau$ reduces the expected value of $p_{_{n,\omega}}$.

\subsection*{Appendix G. Proof of Theorem~\ref{thm:MAD_mean}}\label{appd:G}
When $n$ is large, the condition $(\theta,\rho, \tau)\in \Omega_{\omega,\delta,n}$ implies $\Delta>0$, based on Lemma \ref{lm:solvethetarho_delta}, where $\Delta_1>0$, and $\Delta_3<0$. 
Therefore, as shown in the proof of Lemma \ref{lm:solvethetarho_delta}, both $\theta+s\to\infty$ and $\rho+n-s\to\infty$ as $n\to\infty$.

Applying Stirling's formula, Johnson et al. (1995, p.~219) derive the following result for the ratio of the variance to the squared MAD around the mean:
$$
\lim\limits_{\theta+s\to\infty, \rho+n-s \to\infty}
\frac{\left(\mathbb{E} \left [ \left | p_{_{n,\omega}} - \mathbb{E} (p_{_{n,\omega}}) \right | \right ] \right )^2}{ \mathrm{VAR} (p_{_{n,\omega}})}
= \frac{2}{\pi}.
$$
Thus, minimizing the MAD is equivalent to minimizing the variance of $p_{_{n,\omega}}$ when $n$ is large. 
By Theorem \ref{thm:zero_variance}, this completes the proof of Theorem \ref{thm:MAD_mean}.

\subsection*{Appendix H. Proof of Theorem~\ref{thm:compare_gamma_lambda}}\label{appd:H}
Suppose individual $i$ uniformly outperforms individual $j$, and both have equal employment opportunity.
The expected marginal gain for $j$, denoted $\gamma_j[v]$, is:
$$
\resizebox{.9\textwidth}{!}{$
\begin{array}{rcl}
&&
\sum\limits_{T\subseteq \mathbb{N}: j\in T}
\mathbb{P} (\mathbf{S}=T) [v(T) - v(T\setminus \overline{j})] \\
		
&=&
\sum\limits_{T\subseteq \mathbb{N}: j\in T, i\in T}
\mathbb{P} (\mathbf{S}=T) [v(T) - v(T\setminus \overline{j})] 
+
\sum\limits_{T\subseteq \mathbb{N}: j\in T, i\not \in T}
\mathbb{P} (\mathbf{S}=T) [v(T) - v(T\setminus \overline{j})] \\
		
&\stackrel{Z=T\setminus \overline{j}}{=}&
\sum\limits_{T\subseteq \mathbb{N}: j\in T, i\in T}
\mathbb{P} (\mathbf{S}=T) [v(T) - v(T\setminus \overline{j})] 
+
\sum\limits_{Z\subseteq \mathbb{N}: j\not \in Z, i\not \in Z}
\mathbb{P} (\mathbf{S}=Z\cup \overline{j}) [v(Z \cup \overline{j}) - v(Z)] \\
		
&=&
\sum\limits_{T\subseteq \mathbb{N}: j\in T, i\in T}
\mathbb{P} (\mathbf{S}=T) [v(T) - v(T\setminus \overline{j})] 
+
\sum\limits_{Z\subseteq \mathbb{N}: j\not \in Z, i\not \in Z}
\mathbb{P} (\mathbf{S}=Z\cup \overline{i}) [v(Z \cup \overline{j}) - v(Z)]  \\
		
&\le&
\sum\limits_{T\subseteq \mathbb{N}: j\in T, i\in T}
\mathbb{P} (\mathbf{S}=T) [v(T) - v(T\setminus \overline{i})]
+
\sum\limits_{Z\subseteq \mathbb{N}: j\not \in Z, i\not \in Z}
\mathbb{P} (\mathbf{S}=Z\cup \overline{i}) [v(Z \cup \overline{i}) - v(Z)]\\
		
&\stackrel{T =Z\cup \overline{i} }{=}&
\sum\limits_{T\subseteq \mathbb{N}: j\in T, i\in T}
\mathbb{P} (\mathbf{S}=T) [v(T) - v(T\setminus \overline{i})]
+
\sum\limits_{T\subseteq \mathbb{N}: j\not \in T, i\in T}
\mathbb{P} (\mathbf{S}=T) [v(T) - v(T\setminus \overline{i})] \\
		
&=&
\sum\limits_{T\subseteq \mathbb{N}: i\in T}
\mathbb{P} (\mathbf{S}=T) [v(T) - v(T\setminus \overline{i})] 
\ = \
\gamma_i[v].
\end{array}
$}
$$

Notably, the beta-binomial distribution from Equation~(\ref{eq:probabilitydensity}) is not required for this result. 
Moreover, EEO is not necessary for all other players in $\mathbb{N}$, except for the condition:
$$
\mathbb{P} (\mathbf{S}=Z\cup \overline{i})=\mathbb{P} (\mathbf{S}=Z\cup \overline{j})
\quad \mathrm{for}\ \mathrm{all}\quad 
Z\subseteq \mathbb{N} \setminus \overline{i} \setminus \overline{j}.
$$
This condition ensures that individuals $i$ and $j$ have an equal chance of being hired when both are unemployed.

Similarly, individual $j$'s expected marginal loss, denoted $\lambda_j[v]$, is:
$$
\resizebox{.9\textwidth}{!}{$
\begin{array}{rcl}
&&
\sum\limits_{T\subseteq \mathbb{N}: j\not \in T}
\mathbb{P} (\mathbf{S}=T) [v(T\cup \overline{j}) - v(T)] \\
		
&=&
\sum\limits_{T\subseteq \mathbb{N}: j\not\in T, i\in T}
\mathbb{P} (\mathbf{S}=T) [v(T \cup \overline{j}) - v(T)] 
+
\sum\limits_{T\subseteq \mathbb{N}: j\not\in T, i\not \in T}
\mathbb{P} (\mathbf{S}=T) [v(T\cup \overline{j}) - v(T)] \\
		
&\stackrel{Z=T\cup \overline{j}}{=}&
\sum\limits_{Z\subseteq \mathbb{N}: j\in Z, i\in Z}
\mathbb{P} (\mathbf{S}=Z \setminus \overline{j}) [v(Z) - v(Z\setminus \overline{j})] 
+
\sum\limits_{T\subseteq \mathbb{N}: j\not\in T, i\not \in T}
\mathbb{P} (\mathbf{S}=T) [v(T\cup \overline{j}) - v(T)] \\
		
&=&
\sum\limits_{Z\subseteq \mathbb{N}: j\in Z, i\in Z}
\mathbb{P} (\mathbf{S}=Z \setminus \overline{i}) [v(Z) - v(Z\setminus \overline{j})]
+
\sum\limits_{T\subseteq \mathbb{N}: j\not\in T, i\not \in T}
\mathbb{P} (\mathbf{S}=T) [v(T\cup \overline{j}) - v(T)] \\
		
&\le&
\sum\limits_{Z\subseteq \mathbb{N}: j\in Z, i\in Z}
\mathbb{P} (\mathbf{S}=Z \setminus \overline{i}) [v(Z) - v(Z\setminus \overline{i})] 
+
\sum\limits_{T\subseteq \mathbb{N}: j\not\in T, i\not \in T}
\mathbb{P} (\mathbf{S}=T) [v(T\cup \overline{i}) - v(T)] \hspace{1.5cm}\\
		
&\stackrel{T =Z\setminus \overline{i} }{=}&
\sum\limits_{T\subseteq \mathbb{N}: j\in T, i\not \in T}
\mathbb{P} (\mathbf{S}=T) [v(T \cup \overline{i}) - v(T)] 
+
\sum\limits_{T\subseteq \mathbb{N}: j\not\in T, i\not \in T}
\mathbb{P} (\mathbf{S}=T) [v(T\cup \overline{i}) - v(T)] \\
	
&=&
\sum\limits_{T\subseteq \mathbb{N}: i\not \in T}
\mathbb{P} (\mathbf{S}=T) [v(T\cup \overline{i}) - v(T)] 
\ = \
\lambda_i[v].
\end{array}
$}
$$

In these arguments, we use the identity:
$$
\mathbb{P} (\mathbf{S}=Z\setminus \overline{i})=\mathbb{P} (\mathbf{S}=Z\setminus \overline{j})
$$
for any subset $Z\subseteq \mathbb{N}$ such that $i,j \in Z$.
This identity ensures that individuals $i$ and $j$ have an equal chance of being laid off when both are employed.

\subsection*{Appendix I. Proof of Theorem~\ref{thm:equality_of_outcome}}\label{appd:I}
We allocate $(1-\tau) v(\mathbf{S})$ to the subset $\mathbf{S}$ as employment welfare, and $(\tau-\delta)v(\mathbf{S})$ to the subset $\mathbb{N}\setminus \mathbf{S}$ as unemployment benefits.
When the employed individuals are symmetric in $v$ and have EEO, each employed person receives: 
$$
\frac{(1-\tau) v(\mathbf{S})}{s}  = \frac{(1-\tau) v(\mathbf{S})}{n \omega}
$$ 
as their employment welfare, according to Theorem~\ref{thm:compare_gamma_lambda}.

Similarly, when all unemployed individuals are symmetric in $v$ and have EEO, each receives:
$$
\frac{(\tau-\delta) v(\mathbf{S})}{n - s}  = \frac{(\tau-\delta) v(\mathbf{S})}{n (1-\omega)}
$$ 
as their unemployment benefits. 

Therefore, the condition:
$$
\frac{(1-\tau) v(\mathbf{S})}{s} = \frac{(\tau-\delta) v(\mathbf{S})}{n - s} 
$$
is equivalent to:
$$
\frac{1-\tau }{\omega } = \frac{\tau-\delta}{1 - \omega}.
$$
Solving this equation yields:
$$
\tau = 1 - \omega +\delta\omega.
$$

\subsection*{Appendix J. Proof of Corollary~\ref{cr:equality_of_outcome}}\label{appd:J}
By Corollary~\ref{cr:minimal_phi_rate}, any $\phi$-rate $\tau\ge 1-\omega+\delta\omega$.
Therefore, we have:
$$
1-\tau\le \omega (1-\delta).
$$ 
Multiplying both sides by $(1-\omega)$, we obtain:
$$
(1-\tau)(1-\omega) 
\le \omega (1-\delta)(1-\omega)\\
=  \omega(1-\omega+\delta\omega-\delta)
\le  \omega (\tau-\delta).
$$
Hence, it follows that:
$$
\frac{1-\tau}{\omega} \le \frac{\tau-\delta}{1-\omega}
$$
and
$$
\frac{(1-\tau) v(\mathbf{S})}{s} \le \frac{(\tau-\delta) v(\mathbf{S})}{n - s}.
$$
The remaining arguments follow directly from the proof of Theorem~\ref{thm:equality_of_outcome} in Appendix~\ref{appd:I}.

\end{document}